\documentclass[12pt]{iopart}
\usepackage{graphicx} 
\usepackage{caption}
\usepackage{comment}
\usepackage{amsmath}
\usepackage{adjustbox}

\usepackage{listings}
\usepackage{lineno}
\usepackage{xcolor}

\newcounter{mysec}

\lstdefinelanguage{Julia}{
  keywords={function, end, if, else, elseif, for, while, break, continue, return, try, catch, finally, begin, module, import, export, struct, mutable, abstract, primitive, type, global, local, const, let, do, quote, macro, where},
  sensitive=true,
  comment=[l]\#,
  morecomment=[s]{\#=}{=\#},
  morestring=[b]',
  morestring=[b]",
  alsodigit={., e, E},
}

\begin{document}
\bibliographystyle{unsrt}

\title[Wavefront shaping simulations with augmented partial factorization]{Wavefront shaping simulations with augmented partial factorization}

\author{Ho-Chun Lin, Zeyu Wang and Chia Wei Hsu*}
\address{Ming Hsieh Department of Electrical and Computer Engineering, University of Southern California, Los Angeles, California 90089, USA}
\ead{cwhsu@usc.edu}
\vspace{10pt}
\begin{abstract}
Wavefront shaping can tailor multipath interference to control multiple scattering of waves in complex optical systems. However, full-wave simulations that capture multiple scattering are computationally demanding given the large system size and the large number of input channels. Recently, an ``augmented partial factorization'' (APF) method was proposed to significantly speed-up such full-wave simulations.
In this tutorial, we illustrate how to perform wavefront shaping simulations with the APF method using the open-source frequency-domain electromagnetic scattering solver MESTI. We present the foundational concepts and then walk through four examples: computing the scattering matrix of a slab with random permittivities, open high-transmission channels through disorder, focusing inside disorder with phase conjugation, and reflection matrix computation in a spatial focused-beam basis. The goal is to lower the barrier for researchers to use simulations to explore the rich phenomena enabled by wavefront shaping.
\end{abstract}

\section{Introduction\label{sec:Intro}}

Wavefront shaping~\cite{2007_Vellekoop_OL, 2015_Horstmeyer_NP, 2017_Rotter_RMP,  2022_Yu_Innovation, 2022_Gigan_JoPP,2022_Cao_NP} enables precise control over light propagation and interaction within complex media. Over the past decade, wavefront shaping has found a wide range of applications, such as focusing inside scattering media~\cite{2010_Vellekoop_NP, 2015_Lai_NP,2017_Yu_OE}, endoscopic imaging~\cite{2019_Caravaca-Aguirre_APLP, 2021_Leite_APLP}, tailoring the spectrum~\cite{2012_Small_OL, 2021_Gun_SR}, polarization~\cite{2012_Guan_OL, 2017_Aguiar_SA}, and resolution~\cite{PRL_Putten_2011, 2013_Mahalati_OE}, and enhancing transmission~\cite{2013_Yu_PRL, 2014_Popoff_PRL, 2016_Sarma_PRL, 2017_Hsu_Nphys}, energy deposition~\cite{2022_Bender_Nphys}, and remission~\cite{2022_Bender_PNAS}. An important application is its use for high-resolution imaging deep inside biological tissue~\cite{2020_Yoon_NRP, 2020_Badon_SA, 2023_Zhang_arXiv}. Wavefront shaping can also be used to reconfigure scattering media into functional components like polarization-selective elements~\cite{2018_Xiong_LSA} and linear operators~\cite{2019_Matthes_Optica}, broadening their applications in optical communications~\cite{2021_Ruan_NC} and optical computing~\cite{2022_Gigan_NP}. 

In an experiment, it is challenging (oftentimes impossible) to have access to all input and output channels of the system. Additionally, the internal structure of the complex medium is typically unknown and cannot be readily modified. Numerical simulations provide complete access to the system and the ability to tailor it. It can also save time and resources to test an idea with simulations before performing the experiments. However, wavefront shaping is usually carried out in large-scale multi-channel systems that couple numerous degrees of freedom, and modeling it requires solving one scattering problem for every input channel of interest. To capture the multiple scattering in these complex systems, each scattering problem requires a full-wave solution of Maxwell's equations. The total computation cost often becomes prohibitively high. We recently proposed an ``augmented partial factorization'' (APF) method~\cite{2022_Lin_NCS} to significantly speed up such computations by directly computing the scattering matrix of interest. 

{{In this tutorial, we illustrate how to perform wavefront shaping simulations with the APF method using our open-source frequency-domain electromagnetic scattering solver Maxwell's Equations Solver with Thousands of Inputs (MESTI)~\cite{MESTI_m, MESTI_jl}. MESTI has two programming language interfaces: MESTI.m~\cite{MESTI_m}, which is based on Matlab, and MESTI.jl~\cite{MESTI_jl}, written with the Julia language. While both interfaces share similar syntax and usage, MESTI.jl offers more features, such as the support for full 3D vectorial systems, MPI distributed parallelization, built-in subpixel smoothing, and single-precision arithmetic. In this tutorial, we use MESTI{{.jl (hereafter referred to simply as MESTI)}} and provide four examples to cover the different aspects of common wavefront shaping simulations.}} The full scripts for these examples are provided on the MESTI.jl GitHub repository~\cite{MESTI_jl}. The aim of these examples is to serve as the starting point for researchers to simulate wavefront shaping in a wide range of systems.

\section{Scattering matrix and augmented partial factorization (APF)\label{sec:B_and_C_and_APF}}

\subsection{Maxwell's equations}

We start with time-harmonic electromagnetic waves at frequency $\omega$ (with $e^{-i{\omega}t}$ dependence on time $t$). The frequency-domain Maxwell’s equations for a nonmagnetic system can be written as~{{\cite{Chew1994_JAP, 2013_Shin_phD_thesis}}
\begin{equation}
\left[- \frac{\omega^2}{c^2} {\overline{\overline{\varepsilon}}}_{\rm r}({\bf r}) + \nabla \times \nabla \times\right]{\bf E}({\bf r}) = i\omega\mu_0{\bf J}({\bf r}) \equiv {\bf b}({\bf r}),
\label{eq:Maxwell_eq}
\end{equation} 
where ${\overline{\overline{\varepsilon}}}_{\rm r}$ is the relative permittivity tensor defining its structure, and ${\bf E}$ and ${\bf J}$  are the electric field profile and electric current source density. The right-hand side ${\bf b}({\bf r})$ is what we refer to as the ``source profile.''

The MESTI software and the APF method both apply to any dimension, but for pedagogical purposes in this tutorial, in the following we consider transverse-magnetic (TM) field $E_x(y,z)$ in 2D systems with no $x$ dependence, for which Eq.~(\ref{eq:Maxwell_eq}) simplifies to 
\begin{equation} 
\left[ - \frac{\omega^2}{c^2}\varepsilon_{xx}(y,z) - \frac{\partial^2}{\partial y^2} -\frac{\partial^2}{\partial z^2} \right] E_x(y,z)= i{\omega}\mu_0J_x(y,z) \equiv b(y,z),
\label{eq:2d_tm_eq}
\end{equation} 
where $\varepsilon_{xx}(y,z)$ is the $xx$-component of the permittivity tensor, $E_{x}(y,z)$ is the $x$-component of the electric field, $J_{x}(y,z)$ is the $x$-component of the current density, corresponding to the right-hand side $b(y,z)$.

\subsection{Scattering matrix {\bf S}\label{sec:s_defintion}}
\begin{figure*}
\centering\includegraphics[width=0.8\textwidth]{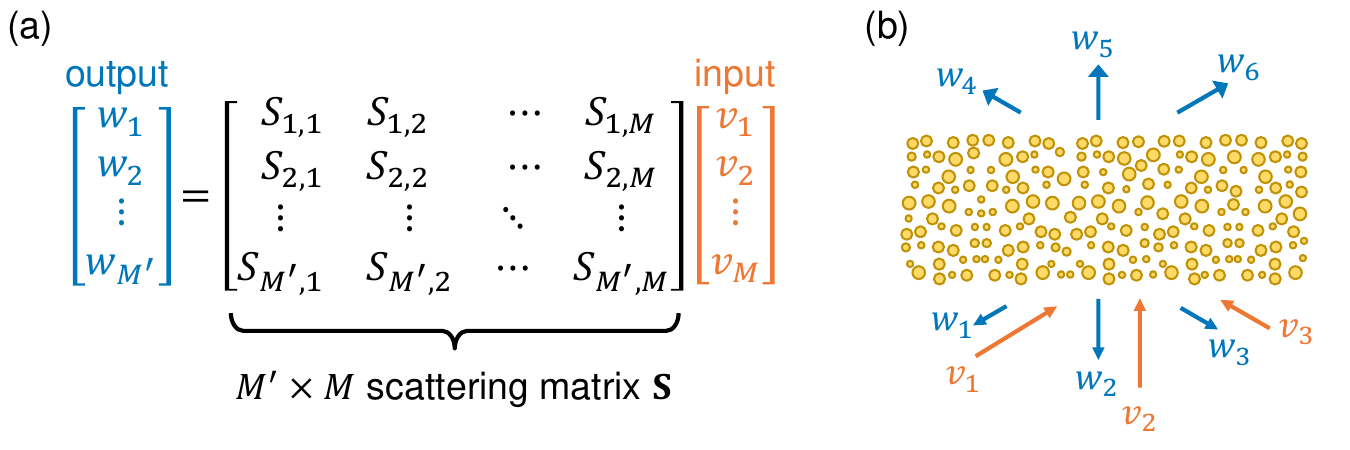}
\caption{{\bf Scattering matrix {\bf S}, input and output.}
{(a)} Scattering matrix {\bf S} relating incident wavefront represented by an input vector $\bf{v}$ in a chosen basis to the resulting output similarly represented by an output vector ${\bf{{w}}}$. The ${w}_n$ and ${v}_m$ give the complex-valued amplitudes in the $n$-th output and $m$-th input basis element. {(b)} Schematic of scattered light (blue arrows) from a disordered structure (yellow circles) from an incident wavefront consisting of multiple input basis elements (orange arrows).}
\label{fig:fig01}
\end{figure*}

Any linear response, including multiple scattering, of complex optical systems can be encapsulated in a scattering matrix {\bf S}. As illustrated in Fig.~\ref{fig:fig01}, any incident wavefront can be represented by an input vector {\bf v} in a chosen basis, and the resulting output (which can be in the near field or far field, in any basis) can be similarly represented as an output vector {\bf w}. The input and output vectors are related by the scattering matrix {\bf S},
\begin{equation}
{{w}_n} = \sum_{m=1}^M {{S_{nm}}{{v}_m}},
 \label{eq:S_eq}
\end{equation} 
where the ${w}_n$ and ${v}_m$ give the complex-valued amplitudes in the $n$-th output and $m$-th input basis element (hereby referred to as ``channels''). We call these ${w}_n$ and ${v}_m$, ``channel amplitude coefficients.'' The $S_{nm}$ element of the $M' \times M$ matrix ${\bf S}$ is the output in the $n$-th channel for an input in the $m$-th channel, where $M$ and $M'$ are {{numbers}} of {{input}} and output channels, respectively. 

Many systems of interest can be naturally divided into two sides, for which
the scattering matrix {\bf S} can be divided into transmission matrices {\bf t} and reflection matrices {\bf r}. Considering input from the low side and output on both sides, the scattering matrix is represented as ${\bf S} = [{\bf r}^{{\rm low}\to{\rm low}}; {\bf t}^{{\rm low}\to{\rm high}}]$. Here, ${\bf r}^{{\rm low}\to{\rm low}}$ denotes the reflection matrix with input from the low side and output on the low side, and ${\bf t}^{{\rm low}\to{\rm high}}$ denotes the transmission matrix with input from the low side and output on the high side.

\subsection{Channels in the homogenous spaces}

For the specific examples below, we consider two-sided systems. The structure consists of an inhomogeneous region with thickness $L$, referred to as the scattering region (inhomogeneous region), sandwiched by homogenous spaces on the two sides (low ($z \le 0$) and high ($z \ge L$) sides),
\begin{equation}
\label{eq:syst}
{\varepsilon _{xx}}(y,z) =
\begin{cases}
\, \varepsilon_{\rm low}, & z \le 0, \\
\, \varepsilon_{xx}(y,z), & 0 < z < L, \\
\, \varepsilon_{\rm high}, & z \ge L.
\end{cases}
\end{equation}
Given the translational symmetry along the $z$-axis, the total field $E_{x}^{(\rm{homo})}(y,z)$ (incident plus scattered) in the homogenous regions ($z \le 0$ or $z \ge L$), {{where ${\varepsilon _{xx}}(y,z)$ = ${\varepsilon _{{\rm{bg}}}}$ is a real-valued constant (either $\varepsilon_{\rm low}$ or $\varepsilon_{\rm high}$),}} can be expressed as a linear combination of propagating and evanescent fields in the two directions: 
\begin{equation}
{E_{x}^{(\rm{homo})}(y,z) = \sum_m v_m^{(+)} E_{x}^{(m,+)}(y,z) + \sum_m v_m^{(-)}E_{x}^{(m,-)}(y,z),}
\end{equation}
with the profile of each ``channel'' being 
\begin{equation}
{E_{x}^{(m,\pm)}(y,z) = \frac{1}{\sqrt{k_z^{(m)}}} u_m(y) e^{ \pm i k_z^{(m)} z }, \quad m \in \mathcal{Z}.}
\label{eq:Ez_m}
\end{equation}
The integer index $m$ labels the channel number, $\pm$ denotes the propagation direction, $k_z^{(m)}$ is the longitudinal wave number, and the prefactor $1/\sqrt{k_z^{(m)}}$ normalizes the amplitude such that the longitudinal flux in $z$ (integrated over $y$) is the same for all propagating channels. The expression for the transverse mode profile $u_m(y)$ depends on the boundary condition along the transverse ($y$) direction. Here, we consider a periodic boundary condition: ${\varepsilon_{xx}(y+W,z) = \varepsilon_{xx}(y,z)}$ and ${E_{x}(y+W,z) = E_{x}(y,z)}$ with a period of $W$, for which the basis mode profile is 
\begin{equation}
u_m(y) = \frac{1}{{\sqrt W }}\exp \left[ {ik_y^{\left( m \right)}\left( {y - {y_0}} \right)} \right], \quad 0 \le y \le W,
\label{eq:u_m}
\end{equation}
where $k_y^{\left( m \right)} =  m\frac{{2\pi }}{W}$ is the transverse wave number and the constant $y_0$ specifies the reference position of the basis. The transverse modes form a complete and orthonormal basis, satisfying 
$\int_0^W u_n^*(y)u_m(y) dy = \delta_{nm}.$
Inserting Eqs.~\eqref{eq:Ez_m}--\eqref{eq:u_m} into Eq.~\eqref{eq:2d_tm_eq} with ${\varepsilon _{xx}}(y,z)$ = ${\varepsilon _{{\rm{bg}}}}$ and no source yields the dispersion relation
\begin{equation}
{\left( {{\omega }/{c}} \right)^2{\varepsilon _{{\rm{bg}}}} = (k_y^{(m)})^2 + (k_z^{(m)})^2.}
\label{eq:kz}
\end{equation}
We choose the $k_z^{(m)} = \sqrt{(\omega/c)^2\varepsilon_{\rm bg} - (k_y^{(m)})^2}$ root. There are infinitely many evanescent channels where $k_z^{(m)}$ takes on imaginary values. There are approximately $2\sqrt{\varepsilon _{\rm{bg}}}W/\lambda_0$ propagating channels, where $k_z^{(m)}$ is real-valued and $\lambda_0 = 2\pi c/\omega$ is the free-space wavelength. Since, in most scenarios, only the propagating ones are of interest, in the following, we primarily consider inputs and outputs in the propagating channels.

\subsection{Source profile $b$ and the corresponding incident field profile \label{sec:source_profile_b}}

There is an important distinction between the source profile $b(y,z)$ on the right-hand side of Eq.~\eqref{eq:2d_tm_eq} and the incident field profile $E_x^{\rm in}(y,z)$ of interest. Oftentimes, we are interested in an incident field profile $E_x^{\rm in}(y,z)$ of the form in Eq.~\eqref{eq:Ez_m} and the resulting scattered field $E_x^{\rm sca}(y,z)$, such that $E_x^{\rm sca}(y,z)$ satisfies an outgoing boundary condition and that the total field $E_x(y,z) = E_x^{\rm in}(y,z) + E_x^{\rm sca}(y,z)$ is a solution of Eq.~\eqref{eq:2d_tm_eq} without a physical current source $b(y,z)$. But computationally, it is more convenient to find an effective source profile $b(y,z)$ that generates the $E_x^{\rm in}(y,z)$ of interest in the absence of the scattering medium, and then solve Eq.~\eqref{eq:2d_tm_eq} with the scattering medium and the effective source profile $b(y,z)$. Different types of sources can be used, such as volume sources, total-field/scattered-field sources~\cite{2012_Rumpf_PIER, 2013_Oskooi_book_chapter} (which are dipole sources with a thickness of two pixels), and surface sources (with a thickness of one pixel). Since the APF method works most efficiently when we have a sparse source profile~\cite{2022_Lin_NCS}, the latter choice (surface source with single-pixel thickness) is preferred, which we describe below.

A surface source $b(x,y)$ at plane $z=z_{\rm s}$ (called ``source plane'') generates light propagating away from $z=z_{\rm s}$. If we want an incident field in the form of Eq.~\eqref{eq:Ez_m}, the surface source should generate field
\begin{equation}
\label{eq:Ez_inc}
E_x^{(m,{\rm source})}(y,z) = \frac{1}{\sqrt{k_z^{(m)}}}  u_m(y) e^{ i k_z^{(m)} |z-z_{\rm s}|}
\end{equation} 
{{in the absence of the scattering medium. Note that this $E_x^{(m,{\rm source})}$ was denoted as $E_{x0}^{{{\rm out}}}$ in Ref.~\cite{2022_Lin_NCS}.}} By substituting Eq.~(\ref{eq:Ez_inc}) into Eq.~(\ref{eq:2d_tm_eq}) with ${\varepsilon _{xx}}(y,z)$ = ${\varepsilon _{{\rm{bg}}}}$ and integrating it over an infinitesimal slice across $z = z_{\rm s}$, we find the corresponding source profile 
\begin{equation}
\label{eq:source_b}
b_m(y,z) = -2i \sqrt{k_z^{(m)}}  u_m(y) \delta\left( z-z_{\rm s} \right),
\end{equation}
where $\delta\left( z \right)$ is the Dirac delta function. This effective surface source $b_m(y,z)$ generates the outgoing field {{$E_{x}^{(m,{\rm source})}(y,z)$}} in Eq.~(\ref{eq:Ez_inc}). For an incident field on the low side that propagates from $z<0$ to $z>L$, we can put the source plane at $z_{\rm s} =0$. Then, on the $z \ge 0$ side, $E_{x}^{(m,{\rm source})}(y,z)$ propagates toward $+z$ and acts as the incident field profile $E_x^{\rm in}(y,z)$ of interest; on the $z<0$ side, {{$E_{x}^{(m,{\rm source})}(y,z)$}} is different from $E_x^{\rm in}(y,z)$, but that has no consequence since there is no scattering at $z<0$. For an incident field on the high side that propagates from $z>L$ to $z<0$, we can put the source plane at $z_{\rm s} = L$.

Note that aside from the exponential and the delta function in $z$, the surface source $b_m(y, z)$ and the field profile {{$E_{x}^{(m,{\rm source})}(y,z)$}} it generates also differ by a prefactor of $-2ik_z^{(m)}$, which depends on the channel index $m$. We should keep this prefactor in mind when we set up the effective source.

When the incident field $E_x^{\rm in}(y,z)$ of interest is a superposition of multiple channels, we can use linearity to find the effective source profile $b(y,z)$ that will generate the $E_x^{\rm in}(y,z)$ of interest.
For example, given a desired incident field profile on the low side
\begin{equation}
E^{\rm{in}}_{x}(y,z) = \sum_m {v}_m E_{x}^{(m,+)}(y,z),
\label{eq:combinatoin_in}
\end{equation}
we use an effective source profile
\begin{equation}
b(y,z) = \sum_m {v}_m b_m(y,z),
\label{eq:b}
\end{equation}
which generates {{$E^{\rm{source}}_{x}(y,z) = \sum_m {v}_m E_{x}^{(m,{\rm source})}(y,z)$}} in the absence of the scattering medium. This {{$E^{\rm{source}}_{x}(y,z)$}} equals the desired incident field $E^{\rm{in}}_{x}(y,z)$ for $z \ge 0$. The coefficient ${v}_m$ here is the input channel amplitude coefficient introduced earlier in Eq.~(\ref{eq:S_eq}). Similar applies to incidence on the high side. 

The above descriptions apply to the continuous wave equation, Eq.~(\ref{eq:2d_tm_eq}). For the wave equation under finite-difference discretization (used in MESTI), the prefactors change slightly. {{The finite-difference version of the equations and prefactors is detailed in Suppl. Sec. 2 of Ref .~\cite{2022_Lin_NCS}.}}

\subsection{Projection profile c \label{sec:projection_c}}

When we numerically solve Eq.~(\ref{eq:2d_tm_eq}) for the scattering medium with the effective source $b(y,z)$ in Eq.~\eqref{eq:b}, we obtain the total field {{$E_{x}(y,z) = E_{x}^{\rm{source}}(y,z) + E_{x}^{\rm{sca}}(y,z)$}} in the regime where $E_{x}^{\rm{source}} = E_{x}^{\rm{in}}$. Subtracting {{$E_{x}^{\rm{source}}(y,z)$}} from the total field then yields the scattered field $E_{x}^{\rm{sca}}(y,z)$ of interest.

However, the full scattered field $E_{x}^{\rm{sca}}(y,z)$ everywhere in space is rarely needed. Typically, we only need to compute projected quantities of the form
\begin{equation}
\iint c(y,z)E_{x}(y,z) dydz.
\end{equation}

As an example, here we consider projections onto the channels defined earlier in Eq.~\eqref{eq:Ez_m}. For instance, on the low side, we have scattered field
\begin{equation}
E_{x}^{\rm{sca}}(y,z) = \sum_{n} {w}_{n} E_{x}^{(n,-)}(y,z), \quad z \le 0.
\label{eq:E_sca}
\end{equation}
The coefficient ${w}_{n}$ here is the output channel amplitude coefficient introduced in Eq.~(\ref{eq:S_eq}). To obtain the coefficient ${w}_{n}$, we can utilize the orthonormality of the transverse modes in Eq.~\eqref{eq:u_m}, multiply $\sqrt{k_z^{({n})}} u^*_{n}(y)$ on both sides of Eq.~(\ref{eq:E_sca}) evaluated at $z = z_{\rm d} = 0$ (called ``detection plane''), and integrate over the transverse direction
\begin{equation}
{w}_{n} = \int_0^W \sqrt{k_z^{({n})}} u_{n}^*(y)E_{x}^{\rm{sca}}(y,z = z_{\rm d}) dy,
\label{eq:w_n}
\end{equation}
which corresponds to a projection profile of
\begin{equation}
c_n(y,z) = \sqrt{k_z^{({n})}}  u^{*}_{n}(y) \delta\left( z - z_{\rm d}\right).
\label{eq:c_n}
\end{equation}
Aside from the exponential and the delta function in $z$, $c_n(y,z)$ in Eq.~\eqref{eq:c_n} and $E_{x}^{(n,-)}(y,z)$ in Eq.~\eqref{eq:Ez_m} differ by a prefactor of $k_z^{(n)}$, and the transverse profile in $c_n(y,z)$ is conjugated.

Note that the projection applies to the total field {{$E_{x}(y,z) = E_{x}^{\rm{source}}(y,z) + E_{x}^{\rm{sca}}(y,z)$}}.
When computing the transmission matrix, it is indeed the total field (incident plus scattered) that we want to project. But for the reflection matrix, we only want to project the scattered field {{$E_{x}^{\rm{sca}}(y,z)$}}; in this case, we need to subtract the projection of {{$E_{x}^{\rm{source}}(y,z)$}}. This subtraction is given by matrix ${\bf D}$, which we describe in the next section.

\subsection{Computation of the scattering matrix\label{sec:s_matrx}}

To perform numerical computation, we represent Eq.~(\ref{eq:2d_tm_eq}) on a finite basis. After such ``discretization,'' Eq.~(\ref{eq:2d_tm_eq}) becomes 
a system of linear equations,
\begin{equation} 
{\bf{A}}{\bf x}_m={\bf b}_m,
\label{eq:Ax_b}
\end{equation}
where matrix ${\bf A}$ is the discretized version of the Maxwell differential operator $- \frac{\omega^2}{c^2}\varepsilon_{xx}(y,z) - \frac{\partial^2}{\partial y^2} -\frac{\partial^2}{\partial z^2}$, the source profile $b(y,z)$ becomes a column vector ${\bf b}_m$, and its corresponding electric field $E_x(y,z)$ is in a column vector ${\bf x}_m$. The subscript $m$ denotes the linear equations corresponding to the $m$-th input channel and its solution. Solving a scattering problem for $m$-th input channel, equivalently, computes ${\bf A}^{-1}{\bf b}_m$ and does the projection through the projection profiles. 

Wavefront shaping is usually carried out in large-scale multi-channel systems, so our interest extends beyond a single input channel and involves multi-channels (denoted as $m = 1,...,M$, where $M$ represents the number of distinct input channels). For multi-channel systems, the scattering matrix {\bf{S}} can be computed through 
\begin{equation} 
{\bf S}= {\bf C}{\bf A}^{-1}{\bf B} - {\bf D},
\end{equation} 
where the $M$ columns of ``source profiles matrix'' ${\bf B} = [{\bf b}_1, ..., {\bf b}_m, ..., {\bf b}_{M}] $  contains the $M$ source profiles, the $M'$ rows of ``projection profiles matrix'' ${\bf C} = [{\bf c}_1; ...; {\bf c}_n; ...; {\bf c}_{M'}] $ consists of the $M'$ projection profiles, where the row vector $c_n$ represents the discretized version of projection profile $c_n(y,z)$. 

The matrix {\bf D} subtracts the baseline contribution from the incident, which can be computed through ${\bf{D}} = {\bf{C}}{\bf A}_0^{-1}{\bf{B}} - {\bf S}_0$~\cite{2022_Lin_NCS}. Here, ${\bf A}_0$ represents the Maxwell differential operator of a reference system ($e.g.$ homogeneous system) for which the scattering matrix ${\bf S}_0$ is known. For transmission calculation ($i.e.$ source region and detection region are on different sides), ${\bf{D}}$ is typically zero.

Above, we consider differential operator ${\bf A}$, surface source profiles ${\bf B}$, and surface projection profiles ${\bf C}$ for TM polarization in 2D. The corresponding expressions for TE polarization in 2D are similar.
The expressions for full 3D vectorial waves are more involved and will be described in a separate paper~\cite {2024_Lin_prepare}. 

\subsection{Augmented partial factorization \label{sec:APF}}

The APF method~\cite{2022_Lin_NCS} lies in computing the entire ${\bf S}$ in a single shot through the partial factorization of an augmented matrix 
\begin{equation}
\label{eq:K}
{
{\bf K} = \begin{bmatrix}
{\bf A} & {\bf B} \\
{\bf C} & {\bf D}
\end{bmatrix} 
=
\begin{bmatrix}
{\bf L} & {\bf 0} \\
{\bf E} & {\bf I}
\end{bmatrix}
\begin{bmatrix}
{\bf U} & {\bf F} \\
{\bf 0} & {\bf H}
\end{bmatrix},
}
\end{equation}
resulting in its minus Schur complement, {{$-{\bf H} \equiv  {\bf C} {\bf A}^{-1} {\bf B} - {\bf D} = {\bf S}$}}, as illustrated in Fig.~\ref{fig:apf}. {{The factorization is partial since it stops after factorizing the upper left block of {\bf K} into {\bf A} ={\bf L}{\bf U}, where {\bf L} and {\bf U} are lower triangular matrix and upper triangular matrix. We perform a partial factorization on an augmented matrix {\bf K}, so this approach is called ``augmented partial factorization''.}} Without looping over inputs and computing unnecessary full solutions everywhere, the APF method computes the whole {\bf S} in a single shot. The computing time and memory scaling of the APF method is characterized by the number of nonzero elements in the augmented matrix {\bf K}, so the APF method is more computationally efficient in using surface (sparse) sources and projection profiles, as we specified in Sec.~\ref{sec:source_profile_b} and Sec.~\ref{sec:projection_c}. If the number of nonzero elements in matrices {\bf B}, {\bf C}, and {\bf S} is fewer than that in matrix {\bf A}, the computational time and memory usage for the APF method is not influenced by the number of input and output channels. APF method stands out for its proficiency in handling multi-channel electromagnetic problems and extending its utility to many applications, such as scattering matrix tomography~\cite{2023_Wang_arXiv} and inverse design~\cite{2023_Li_ACSP}.

\begin{figure*}
\centering\includegraphics[width=1.0\textwidth]{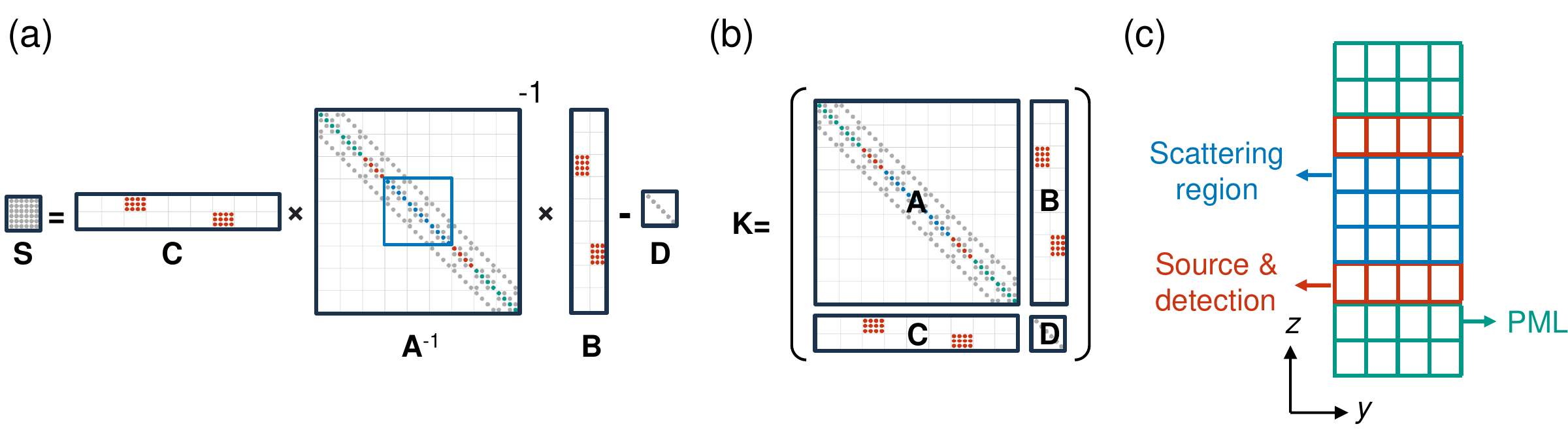}
\caption{{\bf Augmented partial factorization (APF).}
{(a)} Illustration of ${\bf S} = {\bf C} {\bf A}^{-1} {\bf B} - {\bf D}$, where the scattering matrix ${\bf S}$ encapsulates the multi-channel response through the relation ${\bf S} = {\bf C} {\bf A}^{-1} {\bf B} - {\bf D}$. This relationship involves the Maxwell operator matrix ${\bf A}$, source profiles ${\bf B}$, projection profiles ${\bf C}$, and baseline contribution ${\bf D}$. {(b)} Partially factorizing the augmented sparse matrix ${\bf K} = [{\bf A}, {\bf B}; {\bf C}, {\bf D}]$ giving the scattering matrix ${\bf S}$. {(c)} Illustration of the discretization pixels in a 2D system. Each small circle represents a nonzero element of the sparse matrix, with colors indicating its spatial location, as depicted in (a--b). PML: perfectly matched layer.}
\label{fig:apf}
\end{figure*}

\section{Open-source solver MESTI}
\begin{figure*}
\centering\includegraphics[width=1.0\textwidth]{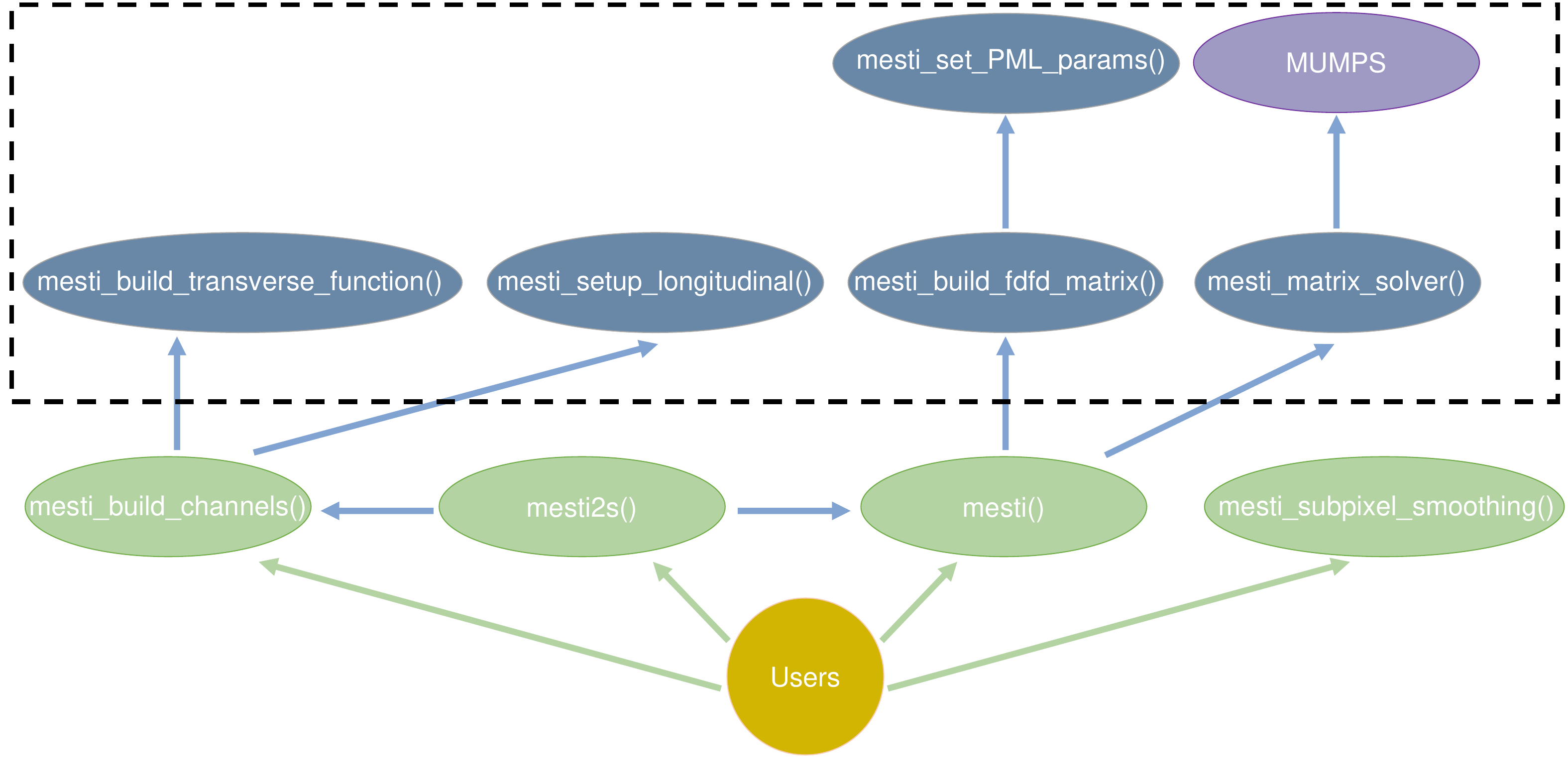}
\caption{{\bf The structure of MESTI.} 
The overall structure and functions of MESTI. Users call four functions (light green nodes) to interact with MESTI. These functions call other internal functions (blue nodes) to construct and solve the systems. MESTI utilizes the MUMPS solver (purple node) to compute the scattering matrix ${\bf S} = {\bf C} {\bf A}^{-1} {\bf B} - {\bf D}$, or full solutions ${\bf X} = {\bf A}^{-1}{\bf B}$. The black dashed box indicates the internal functions within MESTI.}
\label{fig:structure_of_MESTI}
\end{figure*}

We have implemented the APF method in our open-source frequency-domain electromagnetic solver, Maxwell's Equations Solver with Thousands of Inputs (MESTI)~\cite{MESTI_jl}. MESTI solves Maxwell's equations using finite-difference discretization~\cite{2005_FDTD_book,2013_Shin_phD_thesis} on the Yee grid~\cite{1966_Yee_TAP}. It provides a flexible interface, allowing users to compute the scattering matrix ${\bf S} = {\bf C} {\bf A}^{-1} {\bf B} - {\bf D}$ through APF for any relative permittivity profile ${\overline{\overline{\varepsilon}}}_{\rm r}$ (in any dimension), any list of source profiles, and any list of projection profiles. It also supports conventional full solutions ${\bf X} = {\bf A}^{-1}{\bf B}$, where the matrix ${\bf X} = [{\bf x}_1, ..., {\bf x}_m, ..., {\bf x}_{M}]$ contains $M$ field profiles. {{MESTI based on the APF method can handle a large number of inputs simultaneously, so its computational advantage is the most pronounced in multi-input systems like complex media, aperiodic metasurfaces, photonic circuits, and radar cross section computations. One can find benchmarks comparing the APF method with other frequency-domain methods in Ref.~\cite{2022_Lin_NCS}.}}

As described in Sec.~\ref{sec:APF}, a key step of APF is computing the Schur complement of the augmented matrix ${\bf K}$ through a partial factorization. To perform such a partial factorization, MESTI uses the parallel sparse direct solver MUMPS~\cite{2001_Amestoy_SIAM}, {{which is a very efficient code based on the multifrontal method~\cite{1996_Duff_CPC}.} Therefore, to use the APF method in MESTI, the user must also install MUMPS.

MESTI can run on all major operating systems, and detailed installation instructions for Linux, Windows, and Mac are provided in its GitHub repository~\cite{MESTI_jl}. The overall structure of MESTI is illustrated in Fig.~\ref{fig:structure_of_MESTI}. MESTI uses several internal functions to construct the relevant matrices and perform the computations. Users mainly call two functions, $\texttt{mesti()}$ and $\texttt{mesti2s()}$. The first function $\texttt{mesti()}$ provides the most flexibility, and users can specify their own source and projection profiles (the matrices {\bf B} and {\bf C}), in any geometry. The second function \texttt{mesti2s()} (where ``2s'' stands for ``two-sided'') is restricted to two-sided geometries, as described in Eq.~(\ref{eq:syst}), with outgoing boundaries in the longitudinal directions $\pm z$ and closed or periodic boundaries in the transverse directions $(x,y)$, but it is easier to use because it will build the source matrix $\bf{B}$ and the projection matrix $\bf{C}$ for the user. We compare \texttt{mesti()} and \texttt{mesti2s()} and summarize their differences in Tab.~\ref{tab:mesti_mesti2s}. 

\begin{table}[]
\begin{adjustbox}{width=1\textwidth}
\begin{tabular}{|l|l|l|l|l|}
\hline
          & Geometry                                                                                                          & Boundary conditions \& PMLs                                                                                           & Source/projection profiles                                                                                   & Permittivity profiles                                                                                     \\ \hline
\texttt{mesti()}   & General geometry.                                                                                                  & Any.                                                                                                                 &\begin{tabular}[c]{@{}l@{}}$\bullet$ Arbitrary, at any position. \\ $\bullet$ User specified.\end{tabular}                                                                                   & \begin{tabular}[c]{@{}l@{}}Specify ${\overline{\overline{\varepsilon}}}_{\rm r}$ everywhere, including\\ free spaces and PMLs.\end{tabular} \\ \hline
\texttt{mesti2s()} & \begin{tabular}[c]{@{}l@{}}Inhomogeneous region \\ sandwiched by homogenous \\ spaces on the two sides.\end{tabular} & \begin{tabular}[c]{@{}l@{}}Closed/periodic in transverse \\  directions; PMLs in\\ longitudinal direction.\end{tabular} & \begin{tabular}[c]{@{}l@{}}$\bullet$ Superposition of propagating \\ \ \ \ channels in the homogenous space(s). \\ $\bullet$ Automatically built. \end{tabular} & \begin{tabular}[c]{@{}l@{}}Specify  ${\overline{\overline{\varepsilon}}}_{\rm r}$ only in the \\ inhomogeneous region.\end{tabular}           \\ \hline
\end{tabular}
\end{adjustbox}
\caption{{\bf Functions \texttt{mesti()} and \texttt{mesti2s()}.} Comparison between the functions \texttt{mesti()} and \texttt{mesti2s()}. \texttt{mesti()} handles general geometries, arbitrary boundary conditions, and perfectly matched layers (PMLs) and specifies permittivity profiles everywhere. \texttt{mesti2s()} is specialized for inhomogeneous regions sandwiched by homogeneous spaces. \texttt{mesti()} offers versatility across a wider range of scenarios and is able to incorporate the functionalities of \texttt{mesti2s()}.} 
\label{tab:mesti_mesti2s}
\end{table}

\subsection{Maxwell differential operator {\bf A}}
{{Matrix ${\bf A}$ is the discretized form of the Maxwell differential operator, namely $- \frac{\omega^2}{c^2} {\overline{\overline{\varepsilon}}}_{\rm r}({\bf r}) + \nabla \times \nabla \times$ in Eq.~\eqref{eq:Maxwell_eq} in 3D, or $-\frac{\omega^2}{c^2}\varepsilon_{xx}(y,z) - \frac{\partial^2}{\partial y^2} -\frac{\partial^2}{\partial z^2}$ in Eq.~\eqref{eq:2d_tm_eq} for the TM polarization in 2D. In MESTI, the discretization is performed with the finite-difference method~\cite{2005_FDTD_book} on the Yee grid~\cite{1966_Yee_TAP, Chew1994_JAP}, where the space is divided into cubic voxels (or square pixels in 2D as illustrated in Fig.~\ref{fig:apf}c) with size $\Delta x$, the different field components ($i.e.$ $E_x$, $E_y$, $E_z$, $H_x$, $H_y$, and $H_z$) are staggered in space, and the $\nabla \times \nabla \times$ term is discretized with second-order accuracy with $\mathcal{O}(\Delta x^2)$ discretization error. The internal function ${\texttt{mesti\_build\_fdfd\_matrix()}}$ builds matrix ${\bf A}$, but the user needs to specify the discretized form of the relative permittive tensor ${\overline{\overline{\varepsilon}}}_{\rm r}$ and the boundary conditions.

When there is more than one material within a voxel, it can be ambiguous what permittivity one should assign to that voxel. This commonly happens at the interface between two materials. Such a choice also affects the discretization error. A naive assignment may lead to first-order $\mathcal{O}(\Delta x)$ accuracy in the $- \frac{\omega^2}{c^2} {\overline{\overline{\varepsilon}}}_{\rm r}({\bf r})$ term, which spoils the second-order-accurate Yee finite-difference scheme. To resolve this issue, one can assign the discretized permittivity by performing a polarization-dependent tensor average over the materials in each voxel, called ``subpixel smoothing.'' In MESTI, we provide a function ${\texttt{mesti\_subpixel\_smoothing()}}$ that implements the second-order subpixel smoothing scheme described in Ref.~\cite{2006_Farjadpour_OL,2009_Oskooi_OL}.

Another key ingredient in constructing matrix {\bf{A}} is the perfectly matched layer (PML)~\cite{2005_Gedney_book_chapter}, also illustrated in Fig.~\ref{fig:apf}c. In numerical simulations, it is necessary to truncate the computational domain to a finite size with an effective outgoing boundary condition at the truncation interface. The PMLs accomplish this by surrounding the simulation domain with an artificial absorbing layer with minimal reflections back into the domain of interest. The profile of the PML has a great impact on its performance. By default, MESTI uses a resolution-dependent PML profile set by an internal function ${\texttt{mesti\_set\_PML\_params()}}$ that has been optimized~\cite{2024_Wang_arXiv} such that the reflection from the PML and from the truncation interface is negligible for all propagating waves and all evanescent waves even with a thin PML (20 pixels in the examples below) with no homogeneous space added in between the PML and the scattering region. The boundary condition behind the PML (namely, at the truncation interface) is irrelevant since the field is negligibly small there.}}

In MESTI, all the information relevant for constructing the discretized Maxwell operator ${\bf A}$ is specified by the fields of a structure \texttt{syst} in the input arguments of \texttt{mesti()} and \texttt{mesti2s()}. These include the wavelength $\lambda$, discretization voxel size $\Delta x$, the discretized permittivity profile, boundary conditions, and PMLs}.

While most fields of \texttt{syst} are the same across \texttt{mesti()} and \texttt{mesti2s()}, there are some minor differences. Specifically, in \texttt{mesti2s()}, PMLs can only be padded along the longitudinal directions ($\pm z$) (since the transverse directions ($x$ and $y$) have periodic or closed boundaries), while in \texttt{mesti()}, PMLs can be padded in any direction. Also, in \texttt{mesti()}, we specify the discretized permittivity profile everywhere in the computation domain, including inside the PMLs. In \texttt{mesti2s()}, we only specify the discretized permittivity profile in the scattering region [between $0 \le z \le L$ in Eq.~\eqref{eq:syst}], and the constant permittivity on the two homogenous sides (low and high sides) are specified through \texttt{syst.epsilon\_low} and \texttt{syst.epsilon\_high}. 

\subsection{Source and projection profiles, {\bf B} and {\bf C}\label{sec:B_and_C}}
MESTI provides many ways to build the source and projection profiles, {\bf B} and {\bf C}. In \texttt{mesti()}, we have the flexibility to build up customized matrices ${\bf B}$ and ${\bf C}$ by specifying the non-zero elements and their values through cuboids or indices of the spatial locations, as illustrated in Fig.~\ref{fig:b}(a--b). We can also directly call function \texttt{mesti\_build\_channels()} to construct channels and build the {\bf B} and {\bf C} through cuboids. In \texttt{mesti()}, the construction of {\bf B} and {\bf C} provides maximal flexibility based on our interests. 

On the other hand, \texttt{mesti2s()} automatically constructs propagating channels as the basis for the two-sided system. We can use it by either specifying a specific set of (or all) channels we want to use or a superposition of them. They can be done by specifying their channel indices or their channel amplitude coefficients {\bf v} or {\bf w}, as illustrated in Fig.~\ref{fig:b}(c--d). 

\begin{figure*}
\centering\includegraphics[width=1.1\textwidth]{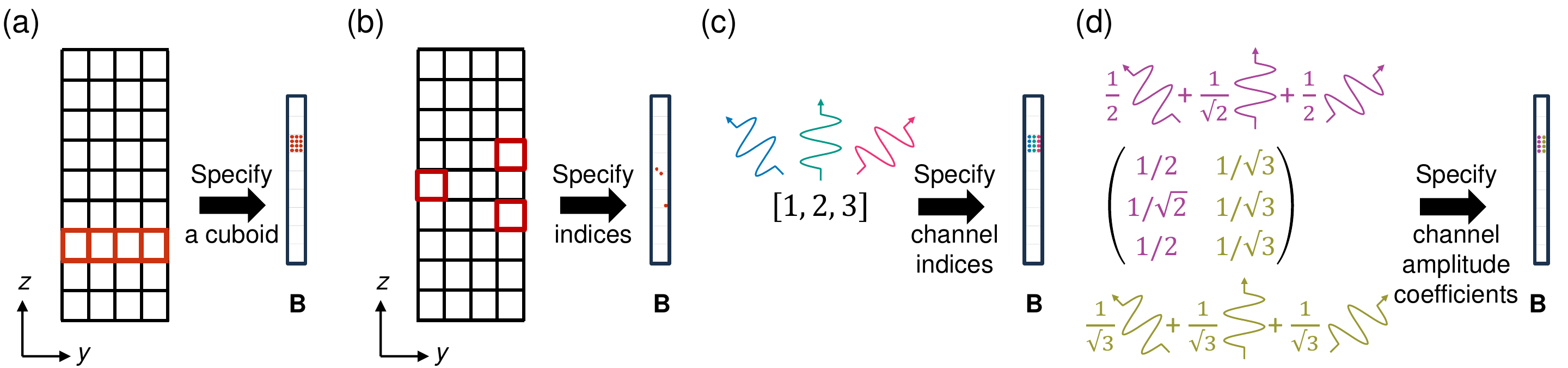}
\caption{{\bf Schematics of specifying source profiles {\bf B}.} (a--d) Specifying source profiles {\bf B}. In general scenarios, the source profiles can be set through (a) cuboids or (b) indices of the simulation domain. Within two-sided geometries, where channels serve as the basis, the source profiles can be specified through {\texttt{mesti2s()}} using either (c) channel indices or (d) channel amplitude coefficients. The same can be applied to projection profiles ${\bf C}$. See the caption of Fig.~\ref{fig:apf} for discretization grids and sparse pattern in {\bf B}.} 
\label{fig:b}
\end{figure*}

\subsection{Baseline matrix {\bf D}\label{sec:matrix_D}}
In \texttt{mesti()}, after preparing ${\bf{B}}$ and ${\bf{C}}$, we can call \texttt{mesti()} to simulate a homogenous reference system ${\bf A}_0$ for which the scattering matrix ${\bf S}_0$ is known, and obtain the baseline matrix {\bf D} through ${\bf{D}} = {\bf{C}}{\bf A}_0^{-1}{\bf{B}} - {\bf S}_0$, as described in Sec.~\ref{sec:s_matrx}.

In \texttt{mesti2s()}, the baseline matrix ${\bf{D}}$ is automatically built based on the inputs and outputs specified by the user, so the user does not need to specify ${\bf D}$. 

\section{Wavefront shaping simulations\label{sec:example}}
In this section, we walk through four examples of using the APF method and the MESTI code: scattering matrix computation in a slab with random permittivities, open high-transmission channels through disorder, focusing inside disorder with phase conjugation, and reflection matrix computation in a spatial focused-beam basis. {{For the first example, we write down the complete Julia script here. For the remaining examples, we walk through part of the Julia script here and provide the full scripts on the MESTI GitHub repository~\cite{MESTI_jl}.}}

\begin{figure*}[ht!]
\centering\includegraphics[width=0.8\textwidth]{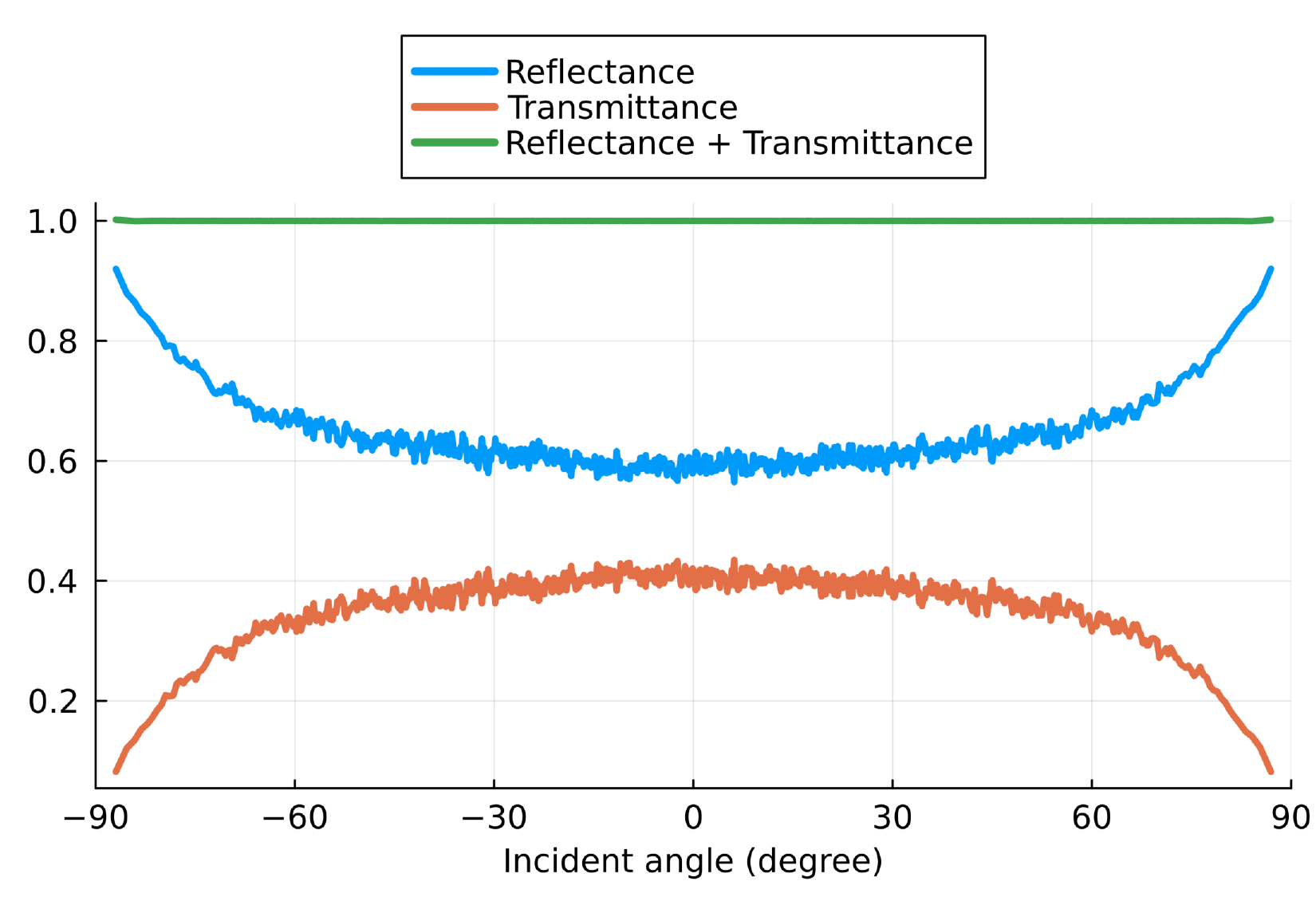}
\caption{{\bf Energy conservation in a disordered slab with a random permittivity at every pixel.}
Around 1,000 propagating channels on each side of the slab. For each incident angle, the curves show the total reflection and the total transmission summed over all outgoing angles, and the sum of reflection and transmission.}
\label{fig:energy_conservation}
\end{figure*}

\subsection{{Warm-up example: scattering matrix of a slab with random permittivities} \label{sec:hellow_world}}

{{We start with a minimal example of using \texttt{mesti2s()} to calculate the scattering matrix {\bf S} with the APF method. We consider a disordered slab with thickness $L = 100\lambda_0$, width $W = 500\lambda_0$ with a periodic transverse boundary, and with $\varepsilon_{xx}$ randomly sampled between 1 and 2.25 at every pixel, where the discretization pixel size is $\Delta x = \lambda_0/15$. 

In this two-sided geometry, users can specify the side field in the arguments {\texttt{input}} and {\texttt{output}}, and {\texttt{mesti2s()}} will automatically prepare the propagating channels as a basis for users based on Eq.~\eqref{eq:source_b} and Eq.~\eqref{eq:c_n}. Here to compute the full scattering matrix \textbf{S}, we set {\texttt{input.side = "low"}} and {\texttt{output.side = "both"}}. It means that we use all propagating channels on the low side as a basis for the input and all propagating channels on both the low and high sides as a basis for the output.

After computing the scattering matrix at wavelength $\lambda_0$, we verify energy conservation by examining the total transmission (summed over the roughly $2W/\lambda_0 = 1,000$ transmitted angles) and the total reflection (summed over all reflected angles) for all incident angles.
The result, shown in Fig.~\ref{fig:energy_conservation}, verifies that \texttt{mesti2s()} adopts the correct normalization of the longitudinal flux for all propagating channels and that the 20 pixels of PML used is effective in realizing an outgoing boundary for all channels. The following is the complete Julia script.}}

\begin{lstlisting}[numbers=none]
using MESTI, Random, Plots

# dimensions of the system, in units of the wavelength lambda_0
dx      = 1/15  # discretization grid size
W       = 500     # width of the disordered slab
L       = 100     # thickness of the disordered slab
ny      = Int(W/dx)  # number of pixel along y-direction
nz      = Int(L/dx)  # number of pixel along z-direction
epsilon_low  = 1.0^2  # permittivity in the low homogenous region
epsilon_high = 1.0^2  # permittivity in the high homogenous region
yBC = "periodic" # boundary condition in y-direction

# the structure syst
syst = Syst() # initialize the structure
syst.epsilon_low  = epsilon_low  # homoegneous permittivity on the low side
syst.epsilon_high = epsilon_high # homoegneous permittivity on the high side
Random.seed!(0) # reproducible random seed
syst.epsilon_xx = 1.0^2 .+ 1.25*rand(ny,nz)  # permittivity profiles in the slab
syst.length_unit  = "lambda_0" # unit of legnth
syst.wavelength = 1 # free space wavelength
syst.dx = dx # discretization grid size
syst.yBC = yBC # boundary condition in y-direction

# PML along the +z and -z directions
pml_npixels = 20 # 20 pixels of PML
pml = PML(pml_npixels)
syst.zPML = [pml]

# the structure input and the structure output
input = channel_type()
output = channel_type()
input.side  = "low"  # input from the low side
output.side = "both" # output to the low and high sides

# mesti2s() is called to compute the scattering matrix S.
S, channels, _ = mesti2s(syst, input, output) # in this setup, S = [r; t]
N_prop_low = channels.low.N_prop # number of propagating channels on the low side
r = S[1:N_prop_low, :] # reflection matrix 
t = S[N_prop_low+1:end, :] # transmission matrix

# plotting
R_list =  reshape(sum(abs2, r, dims=1), :) # reflectance R for all input channels
T_list = reshape(sum(abs2, t, dims=1), :) # transmittance T for all input channels
angle =  atand.(channels.low.kydx_prop./channels.low.kzdx_prop) 
plot(angle, R_list, label="Reflectance R")
plot!(angle, T_list, label="Transmittance T")
plot!(angle, R_list .+ T_list, label="T + R")
xlabel!("Incident angle (degree)")
\end{lstlisting}

\subsection{Open high-transmission channel through disorder\label{sec:open_channel}}

\begin{figure*}
\centering\includegraphics[width=1.0\textwidth]{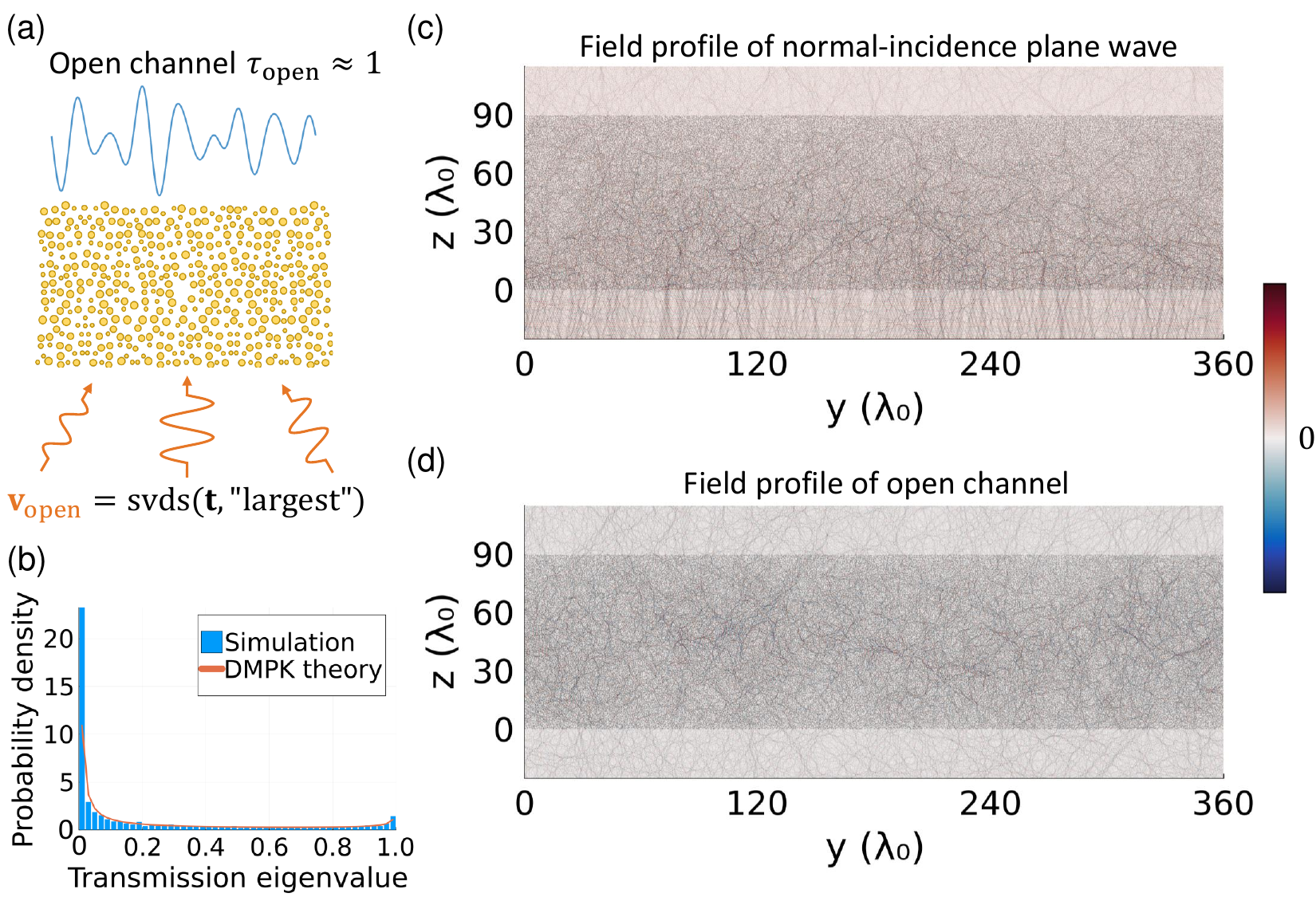}
\caption{{\bf Open channels through disorder.}
{(a)} Illustration of the open high-transmission channel. To identify the open channel with the largest eigenvalue, we perform singular value decomposition (SVD) on the transmission matrix ${\bf t}$ for the disordered system and utilize the right singular vector with the largest singular value as the incident wavefront to achieve high transmission, $\tau_{\rm open} \approx 1$. {{(b) The probability density of the transmission eigenvalue from the simulation and DMPK theory.}} {{{(c--d)}}} The field profiles, ${\rm{Re}}\{E_x\}$, for {{(c)}} the normal incidence propagating channel and {{(d)}} the open high-transmission channel incident from the low side ($z \leq 0$). The open high-transmission channel shows a significantly larger transmission field ($z \geq 90 \lambda_0$) than the normal incidence one. Gray circles represent the scatterers in {{(c--d)}}.}
\label{fig:open_channel}
\end{figure*}

Next, we delve into more wavefront shaping applications using MESTI. In this example, we consider a strongly scattering and opaque (low average transmission) disordered system, compute its transmission matrix {\bf t} using the APF method, and take the information in {\bf t} to realize an almost perfect transmission channel, called the ``open channel~\cite{1984_Dorokhov_SSC,1988_Mello_AnnPhys}.''

{{In a disordered medium, waves can scatter multiple times due to its inhomogeneities, leading to intriguing interference. By tailoring the incident wavefront, one can manipulate this multiple scattering inside a disordered medium and significantly modify the way light travels through the medium, such as its transport. Here, we manipulate the incident wavefront to maximize transmission. To define transmission, we begin by the transmission matrix \textbf{t}, the incident input \textbf{v}, and the transmitted output \textbf{w}. They are related by {\bf w} = {\bf tv}, which only considers transmitted outputs in Eq.~\eqref{eq:S_eq}. The transmitted flux $\langle {\bf w} \vert {\bf w} \rangle $ of the system equals the expectation value $\langle {\bf v} \vert {\bf t}^\dag {\bf t} \vert {\bf v} \rangle $ of the Hermitian matrix ${\bf t}^\dag {\bf t}$. The variational principle guarantees that given an arbitrary incident wavefront \textbf{v} with a normalized incident flux $\langle {\bf v} \vert {\bf v} \rangle$ (here $\langle {\bf v} \vert {\bf v} \rangle = 1$), the maximal and minimal transmitted fluxes of such incident wavefront \textbf{v} are the external eigenvalues of ${\bf t}^\dag {\bf t}$. The eigenvalues of ${\bf t}^\dag {\bf t}$ are called transmission eigenvalues $\{\tau_j\}$~\cite{2017_Rotter_RMP}, where $j$ is the index for the transmission eigenvalue, and the corresponding eigenvectors represent the incident wavefront resulting in such transmission eigenvalue (or transmitted flux). Transmission is defined as the ratio of transmitted flux to incident flux, $\frac{\langle {\bf w} \vert {\bf w} \rangle}{\langle {\bf v} \vert {\bf v} \rangle}$. Here, as the incident flux is normalized to unity, the transmitted flux can represent the transmission for a given incident wavefront ${\bf v}$.}} The incident wavefront ${\bf{v}}_{\rm open}$ that maximizes the transmission results in the most open channel. Here, instead of solving the eigenvectors of ${\bf t}^\dag {\bf t}$, we perform the singular value decomposition (SVD) on ${\bf t}$, which is computationally more efficient (Fig.~\ref{fig:open_channel}(a)). {{The singular values $\{\sigma_j\}$ of {\bf t} are the square roots of the transmission eigenvalues $\{\tau_j\}$, $\sigma_j = \sqrt{\tau_j}$, and the right singular vectors $\{{\bf{v}}_j\}$ of the ${\bf t}$ are the corresponding eigenvectors of ${\bf t}^\dag {\bf t}$~\cite{2017_Rotter_RMP}. The incident wavefront for the most open channel is the right singular vector ${\bf{v}}_{\rm open}$ with the largest singular value $\tau_{\rm open} \approx 1$.}}

{{Here, we show how to compute the transmission matrix using \texttt{mesti2s()}, and then compute the open channel field profile using either \texttt{mesti2s()} or \texttt{mesti()}.}} We consider a scattering medium with width $W = 360\lambda_0$ and thickness $L = 90\lambda_0$, consisting of around 42,000 randomly positioned cylindrical scatterers (refractive index $n$ = 1.2 and radius ranging from 0.2$\lambda_0$ to 0.4$\lambda_0$). We generate a random configuration with non-overlapping cylinders and build its discretized permittivity profile {{${\texttt {epsilon\_xx}}$ in a function {\texttt{build\_epsilon\_disorder()}}}}, which calls ${\texttt{mesti\_subpixel\_smoothing()}}$ to perform subpixel smoothing for all of the cylinders. In the script below, we skip the parts that are the same as in the previous example.

\begin{lstlisting}[numbers=none]
using MESTI, GeometryPrimitives, LinearAlgebra
# including the function to build epsilon_xx for the disordered medium
include("build_epsilon_disorder.jl")

dx      = 1/15; W       = 360; L       = 90 # in unit of lambda_0
r_min   = 0.2; r_max   = 0.4; min_sep = 0.05; number_density = 1.3; rng_seed = 0; 
epsilon_scat = 1.2^2; epsilon_bg   = 1.0^2; build_TM = true

# generating the disorder configuration and build its permittivity profile  
(epsilon_xx, y0_list, z0_list, r0_list, y_Ex, z_Ex) = 
build_epsilon_disorder(W, L, r_min, r_max, min_sep, number_density, 
                       rng_seed, dx, epsilon_scat, epsilon_bg, build_TM)
syst = Syst(); syst.epsilon_xx = epsilon_xx
\end{lstlisting}
After obtaining {\bf{t}} from {\texttt{mesti2s()}}, we subsequently perform SVD on {\bf t} to obtain their singular values $\sigma_j$, transmission eigenvalues $\{\tau_j\}$, and their corresponding right singular vectors ${\bf{v}}_j$.
\begin{lstlisting}[numbers=none]
t, channels, _ = mesti2s(syst, input, output)
_, sigma, v = svd(t)
 \end{lstlisting}

{{In the diffusive regime (thickness $L$ is larger than the transport mean free path), the Dorokhov-Mello-Pereyra-Kumar (DMPK) theory~~\cite{1984_Dorokhov_SSC,1988_Mello_AnnPhys,1994_Nazarov_PRL} predicts that the $\{\tau_j\}$ has a bimodal probability distribution~\cite{1986_Imry_Europhys_Lett,1992_Pendry_PRSL},}}
\begin{equation}
\label{eq:bimodal_distr}
{{{p}\left( \tau \right) = \frac{{\bar \tau}}{{2\tau\sqrt {1 - \tau} }},}}
\end{equation}
{{where ${\bar \tau} = \langle \tau_j \rangle $ is the average transmission. As shown in Fig.~\ref{fig:open_channel}(b), the numerical ${p}\left( \tau \right)$ agrees with the analytic prediction from Eq.~\eqref{eq:bimodal_distr}.
The peak with transmission eigenvalue equal to unity corresponds to open channels.}} 

The incident wavefront of the most open channel ${{\bf v}_{\rm open}}$ corresponds to the right singular vector with the largest singular value. We assign such ${{\bf v}_{\rm open}}$ to {{\texttt input.v\_low}}, which specifies the input channel amplitude coefficients on the low side. The incident wavefront and the corresponding source profile for the most open channel are then constructed using Eq.~\eqref{eq:combinatoin_in} and Eq.~\eqref{eq:b}. We can use {\texttt{mesti2s()}} to compute the full field profile $E_x({\bf r})$ without projection simply by not providing the {\texttt{output}} argument.

\begin{lstlisting}[numbers=none]
# using the right singular vector with the largest  singular value as the input
v_open = v[:, 1] 
input = wavefront(); input.v_low = reshape(v_open,:,1)

opts = Opts()
L_tot = 270 # L = 90 \lambda_0, so extra 90 \lambda_0 on each side for illustration
nz_low  = round(Int,(L_tot-L)/2/dx); nz_high = nz_low
opts.nz_low = nz_low; opts.nz_high = nz_high
Ex, _, _ = mesti2s(syst, input, opts)
\end{lstlisting}
{{Above, we specify an additional argument {\texttt{opts}}, which contains optional parameters. The {\texttt{opts.nz\_low}} and {\texttt{opts.nz\_high}} here request {\texttt{mesti2s()}} to return the total field profile not only inside the scattering medium, but also inside {\texttt{opts.nz\_low}} pixels of homogeneous space on the low side and {\texttt{opts.nz\_high}} pixels on the high side. {\texttt{mesti2s()}} obtains the field profile there by analytically propagating the incident field and the scattered field in the homogeneous spaces, which is a discretized version of the angular spectrum propagation~\cite{2017_Goodman_Book}. Figure~\ref{fig:open_channel}(c) shows the field profile ${\rm{Re}}\{E_x\}$ of the open channel. We also plot the field profile for the normal-incident plane-wave input as a reference (Fig.~\ref{fig:open_channel}(d)).}} The transmission field ($z > 90 \lambda_0$) through the open channel is significantly larger than that of the normal-incidence plane wave. The transmission for this open channel $\tau_{\rm{open}} \approx 1$ is much higher than the average transmission $ \tau_{\rm{ave}} \sim$ 0.2 in the system. 

{{Lastly, for pedagogical purpose, here we also show how to build the source-profile matrix {{\bf B}} and then use {\texttt{mesti()}} to compute the open-channel field profile. We build matrix {\bf B} by specifying its non-zero values through a cuboid as described in Fig.~\ref{fig:b}(a). Also note that in {\texttt{mesti()}}, we need to specify the permittivity profile of the whole computation domain, including the PMLs.

\begin{lstlisting}[numbers=none]
ny_Ex = size(epsilon_xx,1) # number of pixel along y-direction
syst = Syst()
# In mesti(), we provide the whole epsilon_xx including the scattering region, 
# source/detection region, and PML region.
syst.epsilon_xx = cat(epsilon_low*ones(ny_Ex,pml_npixels+1), epsilon_xx, epsilon_high*ones(ny_Ex,pml_npixels+1), dims=2)

syst.length_unit  = "lambda_0"; syst.wavelength = 1; syst.dx = dx; syst.yBC = yBC 
pml = PML(pml_npixels)
# In mesti(), every direction can have PML, so we specify which directions is padded.
pml.direction = "z"
syst.PML = [pml]

Bx = Source_struct()
# We construct mode profile based on Eq. (7).
u_prop_low_Ex = channels.u_x_m(channels.low.kydx_prop)
# We construct effective source profile based on Eq. (10) and Eq. (12).
# nu = sin(kzdx) is the discrete version of the prefactor kz in Eq. (10).
# sqrt_nu_prop is the square root of nu for the propagating channel.
# We take away the prefactor -2i in Eq. (10) and store it in opts.prefactor.
B_Ex_low = u_prop_low_Ex*(channels.low.sqrt_nu_prop.*exp.((-1im*0.5)*channels.low.kzdx_prop).*v_open) 
# Since we set the reference source plane at z = 0, the factor exp((-1im*0.5)*kzdx)) 
# compensates the discrete spatial index l = 0 (corresponding to z = -0.5dx)  and 
# the reference source plane at z = 0. These factors can be ignored if the precise 
# location of the reference plane is not of interest, and here we add it back since 
# we will compare the field result from mesti2s(), which puts the reference plane 
# in this convention. We can see the more in Suppl. Sec. 2B. in Ref 31.

Bx.data = [reshape(B_Ex_low,:,1)] # the nonzero value of the matrix B 

# The position of the source is specified by a rectangle. [m1, l1, w, h] specifies 
# the location and the size of the rectangle. Here, (m1,l1) is the index of the (y,z) 
# coordinate of the smaller-y, smaller-z corner of the rectangle, at the location of 
# Ex(m1,l1). (w,h) is the width and height of the rectangle line source is put on 
# the lower source region, where is one pixel outside PML region.
Bx.pos = [[1,pml_npixels+1,ny_Ex,1]] 
opts = Opts()
# The -2i prefactor of Eq. (10) is multiplied after obtaining Schur complement.
opts.prefactor = -2im
Ex_prime, _ = mesti(syst, [Bx], opts)
\end{lstlisting}
At the end, we verify that the field profile computed by {{{\texttt{mesti()}} is identical to the field profile computed by {\texttt{mesti2s()}} within the scattering medium.}}
\begin{lstlisting}[numbers=none]
# excluding the extra padding and PML region and compare the field profiles
println("Maximum absolute value of field difference between constructing the source matrix B through mesti2s() and constructing by users = ", 
maximum(abs.(Ex[:,nz_low:end-nz_high,:] - Ex_prime[:,pml_npixels+1:end-pml_npixels-1,:])))
\end{lstlisting}
It prints the element-wise maximum absolute difference between the two field calculations, which is zero.

{{While we consider a 2D system here for simplicity, we also can use MESTI to study open channels in 3D~\cite{2024_Lin_arXiv}, and we provide the script on the MESTI GitHub repository~\cite{MESTI_jl}. The field profile computation can use much more memory than scattering matrix calculation due to the storage of the {\bf{L}}{\bf{U}} factors. One can lower the memory usage by storing matrices {\bf{L}} and {\bf{U}} on the disk by setting {\texttt{opts.write\_LU\_factor\_to\_disk}} {\texttt{=}} {\texttt{true}}.}}

\subsection{Focusing inside disorder with phase conjugation\label{sec:phase_conjugation}}
\begin{figure*}
\centering\includegraphics[width=0.85\textwidth]{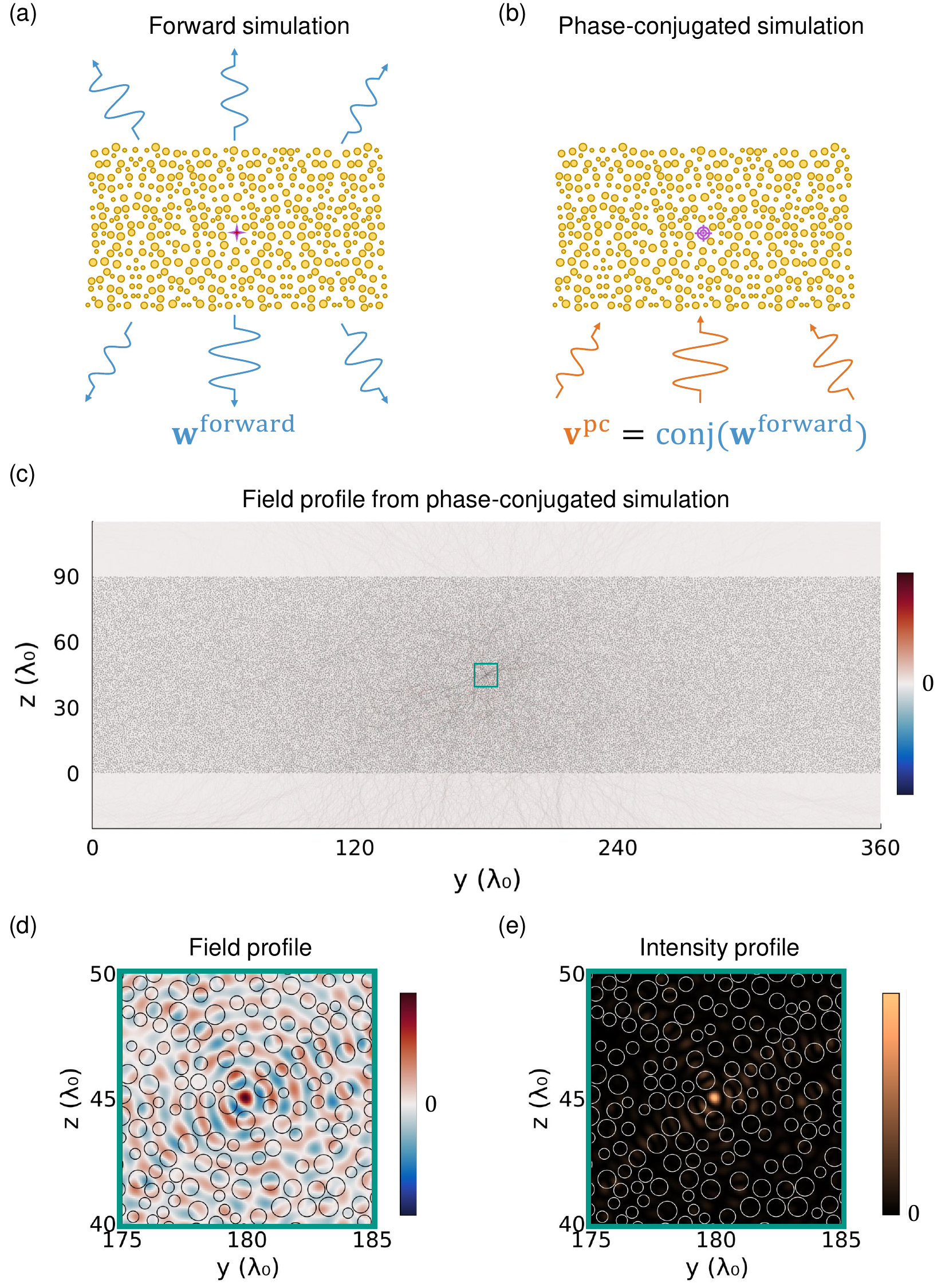}
\caption{{\bf Focusing inside disorder with phase conjugation.}
{{(a--b)}} The phase-conjugated focusing including two sequential simulations. (a) In the forward simulation, the guidestar point-source (four-pointed star) is positioned inside the disordered medium, and the generating field is projected onto the propagating channels on the low side to get the channel amplitude coefficients. (b) In the phase-conjugated simulation,  the channel amplitude coefficients are phase-conjugated and generate the phase-conjugated incident wavefront. The resulting field focuses on the target point (target symbol) inside the disordered medium. {{(c--d)}} The phase-conjugated field profile, ${\rm{Re}}\{E_x\}$, and its zoom-in. {{(e)}} The zoom-in intensity profile, $|E_x|^2$. The green boxes mark the zoom-in regions. Gray circles, black circles, and white circles represent the scatterers in panels (c), (d), and (e), respectively.}
\label{fig:phase_conjugation}
\end{figure*}

Focusing light onto a target point inside a scattering medium is another powerful application of wavefront shaping. In this example, we demonstrate focusing inside a disordered medium with phase conjugation~\cite{2010_Steel_OE, 2015_Tseng_BOE}. This study consists of two simulations: a forward simulation and a phase-conjugated simulation. As illustrated in Fig.~\ref{fig:phase_conjugation}(a), in the forward simulation, a guidestar point-source is placed in the center of the scattering region, which is also our target focusing position later. The resulting fields from the point source are computed and projected onto propagating channels on a detection plane, yielding the outgoing wavefront in terms of the channel amplitude coefficients $\{w^{\rm forward}_m\}$. Subsequently, in the phase-conjugated simulation, we use the phase-conjugated wavefront $\{{\rm conj}(w^{\rm forward}_n)\}$ as the incident wave and compute the resulting field profile to assess how well it focuses onto the target guidestar position, as illustrated in Fig.~\ref{fig:phase_conjugation}(b).

{{We use \texttt{mesti()} in the forward simulation since the source here is not a superposition of propagating channels but a point source~(Tab.~\ref{tab:mesti_mesti2s}). Below we illustrate how to build the source matrix {\bf{B}} with a point source and how to build the projection profiles {\bf{C}} to be used in \texttt{mesti()}. In the phase-conjugated simulation, we construct the wavefront input source, which is a superposition of propagating channels, and we can use either \texttt{mesti()} or \texttt{mesti2s()} in principle.  Since we also want to visualize the field profile in the homogenous spaces on the two sides, we utilize \texttt{mesti2s()}, which can compute the field profile on the two sides when we specify {\texttt{opts.nz\_low}} and {\texttt{opts.nz\_high}}.}} 

We use the same scattering medium as in Sec.~\ref{sec:open_channel}, and the setup for {\texttt {syst}} follows the same procedure. Here we focus on how to set up the input and output. For the forward simulation, we specify the customized source profile {\bf B} through the index (position)  of the point source. {\bf B} contains only a single column and a nonzero element.
\begin{lstlisting}[numbers=none]
# specifying the input (point source in the middle of the disordered)
m0_focus = Int((W/dx)/2); l0_focus = Int((L/dx)/2)+pml_npixels+1
Bx = Source_struct(); Bx.pos = [[m0_focus,l0_focus,1,1]]; Bx.data = [ones(1,1)]
\end{lstlisting}
For the projection profiles, we call {\texttt{mesti\_build\_channels()}} to get information on channels, including the longitudinal wave numbers $k_z^{(n)}$ and the propagating mode profile $u_n(y)$. From them, we can construct the effective projection profile $c_n$ base on Eq~\eqref{eq:c_n}. We collect all $c_n$ and build the projection profiles {\bf C}. 
\begin{lstlisting}[numbers=none]
# channels on the low side
k0dx = 2*pi*syst.wavelength*dx
channels_low = mesti_build_channels(Int(W/dx), yBC, k0dx, epsilon_low)
N_prop_low = channels_low.N_prop # number of propagating channels on the low side 

# projection profiles C on the low side
Cx = Source_struct()
ny_Ex = Int(W/dx)
Cx.pos = [[1,pml_npixels+1,ny_Ex,1]]
# We construct projection profiles based on Eq. (15).
# 0.5 pixel backpropagation indicates that the reference projection plane is at z = 0
C_low = (conj(channels_low.u_x_m(channels_low.kydx_prop))).*reshape(channels_low.sqrt_nu_prop,1,:).*reshape(exp.((-1im*0.5)*channels_low.kzdx_prop),1,:)
Cx.data = [C_low]
\end{lstlisting}
We then compute the channel amplitude ${\bf{w}}^{\rm forward}$  with the APF method by calling \texttt{mesti()}
\begin{lstlisting}[numbers=none]
w, _ = mesti(syst, [Bx], [Cx])
\end{lstlisting}

For the phase-conjugated simulation, we phase-conjugate the previously recorded ${\bf{w}}^{\rm forward}$ and get ${\bf{v}}^{\rm pc}$. From Eq.~\eqref{eq:b}, we specify ${\bf{v}}^{\rm pc}$ in ${\texttt{input.v\_low}}$ to generate a phase-conjugated wavefront to focus on the target point.

\begin{lstlisting}[numbers=none]
input = wavefront()
input.v_low = zeros(ComplexF64, N_prop_low, 1)
# For the phase conjugated input, we have
# conj(coefficient*u) = conj(coefficient)*conj(u) = conj(coefficient)*perm(u(ky)).
# perm() means permute a vector that switches one propagating channel with one
# having a complex-conjugated transverse profile.
input.v_low[:, 1] = conj(w)[channels_low.ind_prop_conj]/norm(w)

opts = Opts()
L_tot = 270
opts.nz_low = round((L_tot-L)/2/dx); opts.nz_high = opts.nz_low

Ex, _, _ = mesti2s(syst, input, opts)
\end{lstlisting}
To illustrate this phase-conjugated focusing, the field profile ${\rm{Re}}\{E_x\}$, its zoom-in, and zoom-in corresponding intensity profile $|E_x|^2$ are shown in Fig~\ref{fig:phase_conjugation}. From them, we can clearly see that this phase-conjugated wavefront successfully focuses on the target point inside the disordered system. 

\subsection{Reflection matrix computation in a spatial focused-beam basis
\label{sec:Gaussian}}

{{At last, we use a more advanced example to illustrate how to build a customized source matrix \textbf{B} and projection matrix \textbf{C} and compute the scattering matrix \textbf{S} (reflection matrix \textbf{r} here) in an arbitrary basis. Here, we consider a Gaussian beam basis. A Gaussian beam is commonly used as an incident wave in experiments. Also, a spatially localized detection in the far-field can be approximated as a projection onto a Gaussian beam.}}

{{To focus on how to build these customized {\bf B} and {\bf C}, we use a simple system with width $W = 20$ \textmu m and thickness $L = 10$ \textmu m consisting of a single dielectric ($n = 1.2$) cylinder with radius = 0.75 \textmu m located at ($y = W/2$, $z = L/2$). Here we also show how to perform subpixel smoothing for this single cylindrical scatter. By defining the domain to perform the subpixel smoothing and the material object (dielectric cylinder here) inside this domain, we can call the function ${\texttt{mesti\_subpixel\_smoothing()}}$ to perform the subpixel smoothing.}}

\begin{lstlisting}[numbers=none]
# We build the relative permittivity profile from the subpixel smoothing.
# domain centering at (W/2, L/2) and with width W and thickness L
W = 20       # width of simulation domain (including PML) (micrometer)
L = 10     # length of simulation domain (including PML) (micrometer)
r_0 = 0.75   # cylinder radius (micrometer)
n_bg   = 1.0 # refractive index of the background
n_scat = 1.2 # refractive index of the cylinder
domain = Cuboid([W/2,L/2], [W,L])
domain_epsilon = n_bg^2 # permittivity of the domain for subpixel smoothing
# object for subpixel smoothing: cylinder located at (W/2, L/2) with radius r_0
object = [Ball([W/2, L/2], r_0)]
object_epsilon = [n_scat^2] # permittivity of the object for subpixel smoothing
yBC = "PEC"; zBC = "PEC" # boundary conditions
# performing subpixel smoothing and obtaining epsilon_xx
epsilon_xx = mesti_subpixel_smoothing(syst.dx, domain, domain_epsilon, object, object_epsilon, yBC, zBC) 
\end{lstlisting}

\begin{figure*}
\centering\includegraphics[width=0.9\textwidth]{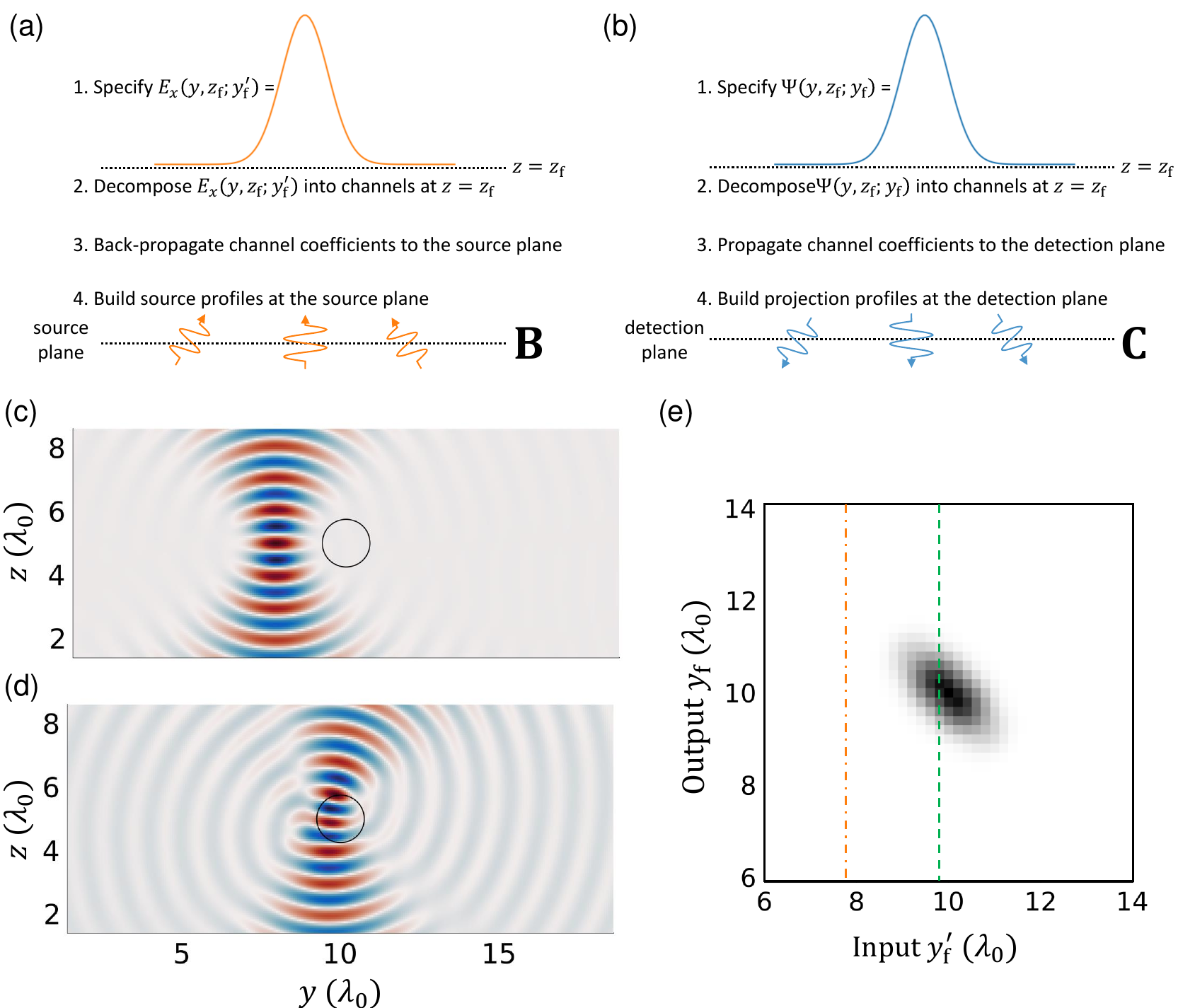}
\caption{{\bf Constructing Gaussian beam basis.}
{(a)} The source profiles {\bf B} for a incident field profile $E_x(y,z_{{\rm f}}; {y_{{\rm f}}^{\prime}})$. To find the source profiles {\bf B} at the source plane $z = z_{{\rm s}}$, the incident Gaussian field profile $E_x(y,z_{{\rm f}}; {y_{{\rm f}}^{\prime}})$ is decomposed into a superposition of channels, and then determine the corresponding channel amplitude coefficients at the source plane by back-propagating them. The source profiles {\bf B} can be built through these channel amplitude coefficients at the source plane.
{(b)} Projetion profiles {\bf C} for profile $\Psi(y,z_{\rm{f}}; y_{\rm{f}})$. Similarly, the projection profiles {\bf C} are built from channel amplitude coefficients at the detection plane $z = z_{{\rm d}}$ through decomposition and propagation. {(c--d)} Incident Gaussian beam with different foci. A single dielectric scatterer is positioned at the center of the simulation domain. The resulting field profiles, ${\rm{Re}}\{E_{x}\}$, are shown for the focus of the incident Gaussian beam (c) away from and (d) near the scatterer. The black circle indicates the scatterer. (e) The corresponding absolute square of the reflection coefficients, $\lvert r_{y_{{\rm f}},{y_{{\rm f}}^{\prime}}} \rvert^2$, with Gaussian beam basis for the single dielectric scatterer system. The orange dashed-dot line corresponds to the incident Gaussian beam in (c), and the green dashed line corresponds to the incident Gaussian beam in (d).}
\label{fig:Gaussian}
\end{figure*}

In the previous examples, we used a periodic boundary condition in the transverse direction $y$ to mimic an infinitely-wide disordered slab.
Here, we consider a finite system, and we put PMLs in all directions (including the transverse direction $y$).
\begin{lstlisting}[numbers=none]
pml_npixels = 20
pml = PML(pml_npixels)
pml.direction = "all" # PMLs in all directions (four sides in 2D)
syst.PML = [pml] 
\end{lstlisting}

\subsubsection{{{Building source profiles {{\bf B}}}}}

{{Here we want incident field profiles $E_x^{\rm in}(y,z)$ to be Gaussian beams focused on $(y, z) = ({y_{{\rm f}}^{\prime}}, z_{\rm{f}})$ in the absence of the scattering medium. The different inputs are Gaussian beams with different transverse focal positions ${y_{{\rm f}}^{\prime}}$, and the source profiles {\bf B} are set up to generate such incident field profiles. As described in Sec.~\ref{sec:APF}, with the APF method, it is more computationally efficient to use a surface source profile. Here, we place the surface sources at $z=z_{\rm{s}} = 0$. Following Sec.~\ref{sec:source_profile_b}, to build the surface source profiles, we can decompose the target incident field profiles $E_x^{\rm in}(y,z)$ into a superposition of channels with different longitudinal wave numbers $k_z^{(m)}$. We specify the transverse profiles of the Gaussian beams at the focal plane $z = z_{\rm{f}}$, back propagate the beams from $z = z_{\rm{f}}$ to the source plane $z=z_{\rm{s}}$, and then build the source profiles ${\bf B}$ there. This process is schematically illustrated in Fig.~\ref{fig:Gaussian}(a). In the following, we walk through the key steps with codes.

We first specify the desired incident Gaussian beam profile at its focal plane $z = z_{\rm{f}}$,}} 
\begin{equation}
E_{x}(y,z_{\rm{f}}; {y_{{\rm f}}^{\prime}}) = \exp\left(-(y - {y_{{\rm f}}^{\prime}})^2/{w_0}^2\right), 
\end{equation}
{{where $w_0 = 2/\pi$ \textmu m is the beam radius.}}
\begin{lstlisting}[numbers=none]
# parameters of the system and input Gaussian beams
wavelength = 1.0 # wavelength (micrometer)
dx = wavelength/15 # discretization grid size (micrometer)
y = dx:dx:(W-dx/2) # grid along y-direction
z = dx:dx:(L-dx/2) # grid along z-direction
ny_Ex = size((dx:dx:(W-dx/2)),1) # number of grid in y-direction
NA = 0.5   # numerical aperture
z_f = L/2  # location of the focus in z (fixed)
y_f_start = 0.3*W # starting location of the focus y_f in y 
y_f_end   = 0.7*W # ending location of the focus y_f in y
dy_f = wavelength/(10*NA)  # spacing of focal spots in y

# parameters of the line sources
l_source = nPML + 1 # index of the source plane
z_source = z[l_source] # location of the source plane

w_0 = wavelength/(pi*NA) # Gaussian beam radius at z = z_f
y_f = y_f_start:dy_f:y_f_end # list of focal positions in y

# size(E_yf) = [ny_Ex, num_of_Gaussian_input]
E_yf = exp.(-(y .- transpose(y_f)).^2/(w_0^2)) 
\end{lstlisting}
{{We call {\texttt{mesti\_build\_channels()}} to obtain propagating mode profiles $u_m(y)$ and the corresponding longitudinal wave numbers $k_z^{(m)}$.}}

\begin{lstlisting}[numbers=none]
n_bg = 1.0 # refractive index of the background

channels = mesti_build_channels(ny_Ex, "PEC", (2*pi/wavelength)*dx, n_bg^2)
# size(u) = [ny_Ex,  num_of_prop_channel]
u = channels.u_x_m(channels.kydx_prop) 
kz = reshape(channels.kzdx_prop/dx, :, 1) # list of propagating wave numbers kz

\end{lstlisting}
{{By decomposing the target incident field profile $E_x(y,z_{\rm f}; {y_{{\rm f}}^{\prime}})$ into a superposition of propagating channels with different longitudinal wave numbers $k_z^{(m)}$, we get the channel amplitude coefficient $v_m^{{\rm{f}},{y_{{\rm f}}^{\prime}}}$ at focal plane $z = z_{{\rm f}}$, We then back-propagate $v_m^{{\rm{f}},{y_{{\rm f}}^{\prime}}}$ from the focal plane to the source plane and accumulate the phase (Eq.~\eqref{eq:Ez_m}) to obtain the amplitude coefficient $v_m^{{\rm{s}},{y_{{\rm f}}^{\prime}}}$ on the source plane $z = z_{{\rm s}}$,}}
\begin{equation}
v_m^{{\rm{s}},{y_{{\rm f}}^{\prime}}} = e^{i k_z (z_{\rm{s}}-z_{\rm f})} v_m^{{\rm{f}},{y_{{\rm f}}^{\prime}}}, \quad
v_m^{{\rm{f}},{y_{{\rm f}}^{\prime}}} = \int_{0}^{W} \sqrt{k_z^{(m)}} u_m^*(y)E_x(y, z_{\rm{f}}; {y_{{\rm f}}^{\prime}}) \,dy.
\label{eq:v_m_decomp}
\end{equation}

\begin{lstlisting}[numbers=none]
# nu = sin(kzdx) is the discrete version of the prefactor kz in Eq. (20).
# sqrt_nu_prop is the square root of nu for the propagating channel.
sqrt_nu_prop = reshape(channels.sqrt_nu_prop, :, 1)
# size(v_f) = [num_of_prop_channel, num_of_Gaussian_input]
v_f = (sqrt_nu_prop.* adjoint(u))*E_yf
# size(v_s) = [num_of_prop_channel, num_of_Gaussian_input]
v_s = exp.(1im*kz*(z_s-z_f)).*v_f 
\end{lstlisting}
{{Based on $v_m^{{\rm{s}},{y_{{\rm f}}^{\prime}}}$, effective source profile $b_{y_{\rm f}}(y,z)$ now can be built through Eqs.~(\ref{eq:source_b})--(\ref{eq:b}),}}
\begin{equation}
b_{y_{\rm f}}(y, z) = -2i\sum_m v_m^{{\rm{s}},{y_{{\rm f}}^{\prime}}} \sqrt{k_z^{(m)}}
u_m(y)\delta(z - z_{\rm{s}}).
\label{eq:v_m_source}
\end{equation}
{{In the codes, we specify the prefactor $-2i$ in {\texttt{opts.prefactor}}. 
This makes it easier for the APF method to utilize the symmetry of the reflection matrix, as we will see later}}. 
\begin{lstlisting}[numbers=none]
opts = Opts()
opts.prefactor = -2im # The -2i prefactor of Eq. (10) is multiplied at the end.

Bx = Source_struct()
Bx.pos = [[1, n_source, ny_Ex, 1]]
# size(Bx.data) = [ny_Ex, num_of_Gaussian_input]
Bx.data = [(u.*transpose(channels.sqrt_nu_prop))*v_s] # Eqs. (10) and (12)
\end{lstlisting}

{{For the source profile above to generate the desired incident field, the source profile needs to be placed in a homogenous space away from PML. 
Therefore, we need to make sure that the surface source is negligibly small inside the PML. Here, we set the system width $W$ to be wider than the scan range of $y_{\rm f}$ plus the beam diameter $2w_0$, such that the relative magnitude of the source {{\bf B}} inside the PML is negligibly small.}}

{{To verify that the source profiles have been set up correctly, we compute the total field profiles using \texttt{mesti()}:}}
\begin{lstlisting}[numbers=none]
# Excluding the PML pixels from the returned field profiles
opts.exclude_PML_in_field_profiles = true

field_profiles, _ = mesti(syst, [Bx], opts)
\end{lstlisting}
{{As shown in Fig.~\ref{fig:Gaussian}(c), when the Gaussian beam is away from the cylinder, the total field is a pristine Gaussian beam. When the Gaussian beam is close to the cylinder, it undergoes scattering (Fig.~\ref{fig:Gaussian}(d)).}}

\subsubsection{{{Building projection profiles {{\bf C}}}}}
{{In the next step, we build the projection profiles {\bf{C}}. The projection profiles need to be chosen to model the desired detection scheme. Here, we assume that the reflected wave is captured by a detector positioned at a plane that images the focal plane at $z = z_{{\rm f}}$ in the absence of scattering. We approximate the detection by taking the field overlap with a focused Gaussian spot,}}
\begin{equation}
\label{eq:Gaussian}
\Psi(y,z = z_{\rm{f}}; y_{\rm{f}}) =\exp\left(-(y - y_{\rm{f}})^2/w_0^2\right),
\end{equation}
{{with a transverse focal position ${y_{{\rm f}}^{\prime}}$.}}
{{In the absence of scattering, such a detection can be performed numerically by projecting the scattered field $E_{x}^{\rm{sca}}(y,z)$ at $z = z_{{\rm f}}$ onto Eq.~\eqref{eq:Gaussian}. However, light would experience scattering during the propagation from $z = z_{{\rm f}}$ to the far-field, where the detector is placed. Therefore, we need to perform the projection at a sufficient distance where there is no more scattering. Here we choose to perform the projection at detection plane $z = z_{{\rm d}} = 0$. So, we need to find out the transverse profile of the Gaussian beam at a distance of $z_d - z_f$ from the focal plane.

Since it is easier to propagate waves through propagating channels on an angular basis in homogenous space, as shown in Fig.~\ref{fig:Gaussian}(b), we decompose the Gaussian beam onto the propagating channels and propagate them from the focal plane to the detection plane to get its amplitude coefficients $v_n^{{\rm{d}},y_{\rm{f}}}$ at $z = z_{\rm{d}}$,}}
\begin{equation}
v_n^{{\rm{d}},y_{\rm{f}}} = e^{-i {k_z^{(n)}} (z_{\rm{d}}-z_{\rm f})} {\tilde{v}}_n^{{\rm{f}},y_{\rm{f}}}, \quad
{\tilde{v}}_n^{{\rm{f}},y_{\rm{f}}} = \int_{0}^{W} \sqrt{k_z^{(n)}} u_n^*(y)\Psi(y, z = z_{\rm{f}}; y_{\rm f}) \,dy.
\end{equation}
{{We then reconstruct the transverse profile of the Gaussian beam at the detection plane $z = z_{\rm d}$,}}
\begin{equation}
\label{eq:Psi_d}
\Psi(y, z = z_{\rm{d}}; {y_{\rm f}}) = \sum_n v_n^{{\rm{d}},y_{\rm{f}}} \frac{1}{\sqrt{k_z^{(n)}}}
u_n(y).
\end{equation}
{{Now, we recall Eq.~\eqref{eq:c_n} in Sec.~\ref{sec:projection_c}. When we perform projection onto a target channel profile, we need to multiply a prefactor $k_z^{(m)}$ and conjugate the mode profile. Therefore, the projection profile corresponding to $\Psi(y, z = z_{\rm{d}}; {y_{\rm f}})$ in Eq.~\eqref{eq:Psi_d} is}}
\begin{equation}
c_{y_{\rm f}}(y, z) = \sum_n v_n^{{*\rm{d}},y_{\rm{f}}} \sqrt{k_z^{(n)}}
u^*_n(y)\delta(z - z_{\rm{d}}).
\end{equation}

\begin{lstlisting}[numbers=none]
# We set the detection plane on the same plane as the source plane.
Cx = Source_struct()
Cx.pos = Bx.pos

# size(Psi_yf) = [ny_Ex, num_of_Gaussian_input]
Psi_yf = exp.(-(y .- transpose(y_f)).^2/(w_0^2)) 

# We decompose Psi_yf onto the propagating channels  and obtaining v_f_tilde.
# size(v_f_tilde) = [num_of_prop_channel, num_of_Gaussian_output]
v_f_tilde = (sqrt_nu_prop.* adjoint(u))*Psi_yf 

# We propagate the amplitude coefficient from plane z = z_f to plane z = z_d.
# size(v_d) = [num_of_prop_channel, num_of_Gaussian_output]
v_d = exp.(1im*kz*(z_d-z_f)).*v_f_tilde

# We reconstruct Psi_yf at the detection plane z = z_d.
# Psi_yf_zd = (u./transpose(sqrt_nu_prop))*v_d

# We obtain the projection profiles based on Psi_yf_zd.
Cx.data = [adjoint(v_d)*(sqrt_nu_prop.*adjoint(u))]
\end{lstlisting}

{{It turns out---as we expect from reciprocity---that ${\bf C} = {\bf B}^{\rm T}$ when we take the $-2i$ prefactor away from $\bf{B}$:}}
\begin{lstlisting}[numbers=none]
println("max(|C_low-transpose(B_low)|) = $(maximum(abs.(Cx.data[1] - transpose(Bx.data[1]))))")
\end{lstlisting}
{{It prints the element-wise maximum absolute difference between $\bf{C}$ and ${\bf B}^{\rm T}$, which is zero. When matrix {\bf A} is a symmetric matrix and when ${\bf C} = {\bf B}^{\rm T}$, we can utilize the symmetry of the augmented matrix ${\bf K}$ in Eq.~\eqref{eq:K} and save computing resources, as mentioned in Suppl. Sec. 3 in Ref.~\cite{2022_Lin_NCS}. To utilize this symmetry in $\texttt{mesti()}$, we can simply use}}
\begin{lstlisting}[numbers=none]
C = "transpose(B)"
\end{lstlisting}

\subsubsection{{{Building baseline matrix {{\bf D}}}}}
{{As described in Secs.~\ref{sec:projection_c}--\ref{sec:s_matrx}, for the reflection matrix {\bf r} (source and detection planes are on the same sides) calculation, we also need a baseline matrix ${\bf{D}}$, which can be computed by ${\bf r}_0 = {\bf{C}}{\bf A}_0^{-1}{\bf{B}} - {\bf{D}}$.
${\bf{A}}_0$ is the homogeneous-space Maxwell operator with no scatterers, and ${\bf{r}}_0$ is the reflection matrix in the homogeneous space with no scatterers. The reflection matrix ${\bf{r}}_0$ in {{the}} homogeneous system is just a zero matrix, so  we keep all other input arguments and assign the discretized permittivity profile \texttt{syst.epsilon\_xx} to a homogenous one and call \texttt{mesti()} to calculate the matrix ${\bf{D}} = {\bf{C}}{\bf A}_0^{-1}{\bf{B}}$.}}
\begin{lstlisting}[numbers=none]
# For a homogeneous space, the length of the simulation domain does not matter, 
# so we choose a minimal thickness  pml_npixels + 1 + pml_npixels.
# 1 is the thickness for the homogeneous space. 
syst.epsilon_xx = n_bg^2*ones(ny_Ex, pml_npixels + 1 + pml_npixels)
D, _ = mesti(syst, [Bx], C, opts)
\end{lstlisting}
{{Subsequently, replacing the discretized permittivity profile \texttt{syst.epsilon\_xx} with the target system (a single cylinder dielectric in the middle), we call \texttt{mesti()} again and compute the reflection matrix ${\bf{r}} = {\bf{C}}{\bf A}^{-1}{\bf{B}} - {\bf D}$ with the APF method.}}
\begin{lstlisting}[numbers=none]
syst.epsilon_xx = epsilon_xx
r, _ = mesti(syst, [Bx], C, D, opts)
\end{lstlisting}
{{The resulting absolute square of reflection coefficient, $\lvert r_{y_{{\rm f}},y_{{\rm f}}^{\prime}} \rvert^2$ of such basis is shown in Fig.~\ref{fig:Gaussian}(e). There are large reflections when generating Gaussian beams approach the cylinder center, $y = 10$ \textmu m. 

Such a reflection matrix {\bf r} computation with a customized basis can be used to model optical imaging methods such as optical coherence tomography~\cite{2003_Fercher_PPP} and matrix-based imaging methods~\cite{2017_Kang_NC, 2020_Badon_SA, 2023_Lee_NC, 2023_Zhang_arXiv}, as we did in Ref.~\cite{2023_Wang_arXiv}.}}

\section{Outlook\label{sec:outlook}}
APF is a powerful method that enables many studies of multi-channel systems that used to require too much computing resources to carry out. We hope the pedagogical explanations and the walk-through examples here, combined with the open-source MESTI code, can lower the barrier for researchers to adopt the APF method and explore a wide range of studies in multi-channel systems.

\section*{Data and code availability}
The software Maxwell’s Equations Solver with Thousands of Inputs (MESTI), documentation, data, and examples are available on GitHub~\cite{MESTI_jl} under the GPL-3.0 license.

\section*{Acknowledgement}
We thank S. Li, M. Dinh, and M. Torfeh for useful discussions. This work is supported by the National Science Foundation CAREER award (ECCS-2146021) and the Sony Research Award Program.

\section*{References}

\bibliography{main}

\end{document}